 \documentclass[prl,amsmath,amssymb,twocolumn, showpacs, superscriptaddress,10pt]{revtex4-1}

\usepackage{amsmath}
\usepackage{hyperref}
\usepackage{graphicx}
\usepackage{amsfonts}
\usepackage{amsthm}
\usepackage{cases}
\usepackage{bm}

\usepackage{color}
\definecolor{Blue}{rgb}{0.00, 0.00, 1.00}
\definecolor{Red}{rgb}{1.00, 0.00, 0.00}

\hypersetup{
    colorlinks=true,       
    linkcolor=red,          
    citecolor=blue,        
    filecolor=magenta,      
    urlcolor=cyan           
}

\newcommand{\nn}{\nonumber}
\newcommand{\be}{\begin{equation}}
\newcommand{\ee}{\end{equation}}
\newcommand{\bea}{\begin{eqnarray}}
\newcommand{\eea}{\end{eqnarray}}

\newcommand{\beq}{\begin{equation}}
\newcommand{\eeq}{\end{equation}}
\newcommand{\beqn}{\begin{eqnarray}}
\newcommand{\eeqn}{\end{eqnarray}}

\DeclareMathOperator{\J}{J}
\DeclareMathOperator{\Det}{Det}

\def\q{\frac{\hbar^2}{2m}}

\newcommand{\abs}[1]{\ensuremath{\left| #1 \right|}}

\begin{document}

\title{Statistics of fermions in a $d$-dimensional box near a hard wall}

\author{Bertrand Lacroix-A-Chez-Toine}
\affiliation{LPTMS, CNRS, Univ. Paris-Sud, Universit\'e Paris-Saclay, 91405 Orsay, France}
\author{Pierre Le Doussal}
\affiliation{CNRS-Laboratoire de Physique Th\'eorique de l'Ecole Normale Sup\'erieure, 24 rue Lhomond, 75231 Paris Cedex, France}
\author{Satya N. \surname{Majumdar}}
\affiliation{LPTMS, CNRS, Univ. Paris-Sud, Universit\'e Paris-Saclay, 91405 Orsay, France}
\author{Gr\'egory \surname{Schehr}}
\affiliation{LPTMS, CNRS, Univ. Paris-Sud, Universit\'e Paris-Saclay, 91405 Orsay, France}

\date{\today}

\begin{abstract}
We study $N$ noninteracting fermions in a domain bounded by a hard wall potential
in $d \geq 1$ dimensions. We show that for large $N$, the correlations at the edge of the Fermi gas
(near the wall) at zero temperature are described by a universal
kernel, different from the universal edge kernel valid for smooth potentials. 
We compute this $d$ dimensional hard edge kernel exactly for a spherical domain and argue, using 
a generalized method of images, that it holds close to any sufficiently smooth boundary.
As an application we compute the quantum statistics of the position of the fermion closest
to the wall. Our results are then extended in several directions, including
non-smooth boundaries such as a wedge, and also to finite temperature. 
\end{abstract}



\maketitle

Noninteracting Fermi gas in a confining trap is a subject of great current interest, both theoretically~\cite{GPS08} and in cold atom experiments~\cite{BDZ08}. The trap introduces a soft edge to the Fermi gas where the average density vanishes at zero temperature ($T=0$). Near the edge, the quantum and thermal fluctuations play an important role \cite{Kohn}. 
For a harmonic trap in one-dimension ($d=1$) at $T=0$, the positions of the fermions are in one-to-one correspondence with the eigenvalues of the Gaussian Unitary Ensemble (GUE) of Random Matrix Theory (RMT)~\cite{CMV2011,CMV2011a,Eis2013,marino_prl}. Consequently, the quantum correlations at the edge of the trap are described by the fluctuations of the largest eigenvalues of the GUE~\cite{us_finiteT}. Furthermore, it was shown recently that these edge correlation functions for the harmonic trap
are universal with respect to a large class of smooth confining potentials, e.g. $V(x) \sim |{x}|^p$ with $p>0$. Similarly, the edge correlations for the harmonic trap were shown to be universal in $d>1$ and $T>0$ for smooth potentials~\cite{us_finiteT,DPMS:2015,fermions_review}. It is natural to ask what type of trap potentials lead to edge physics that deviates from this universal description? This is particularly relevant as the current experimental techniques are able to design traps of varying shapes \cite{BDZ08,Zwi2017}. The simplest and perhaps the most natural candidate is a ``box'' with hard wall potential, a standard subject in basic quantum mechanics. In this Letter we show that fermions near the hard wall of a $d$-dimensional box have  universal correlations, e.g., independent of the shape of the box, which are rather different from their counterparts in smooth potentials.


\begin{figure}[ht]
\includegraphics[width=0.8\linewidth]{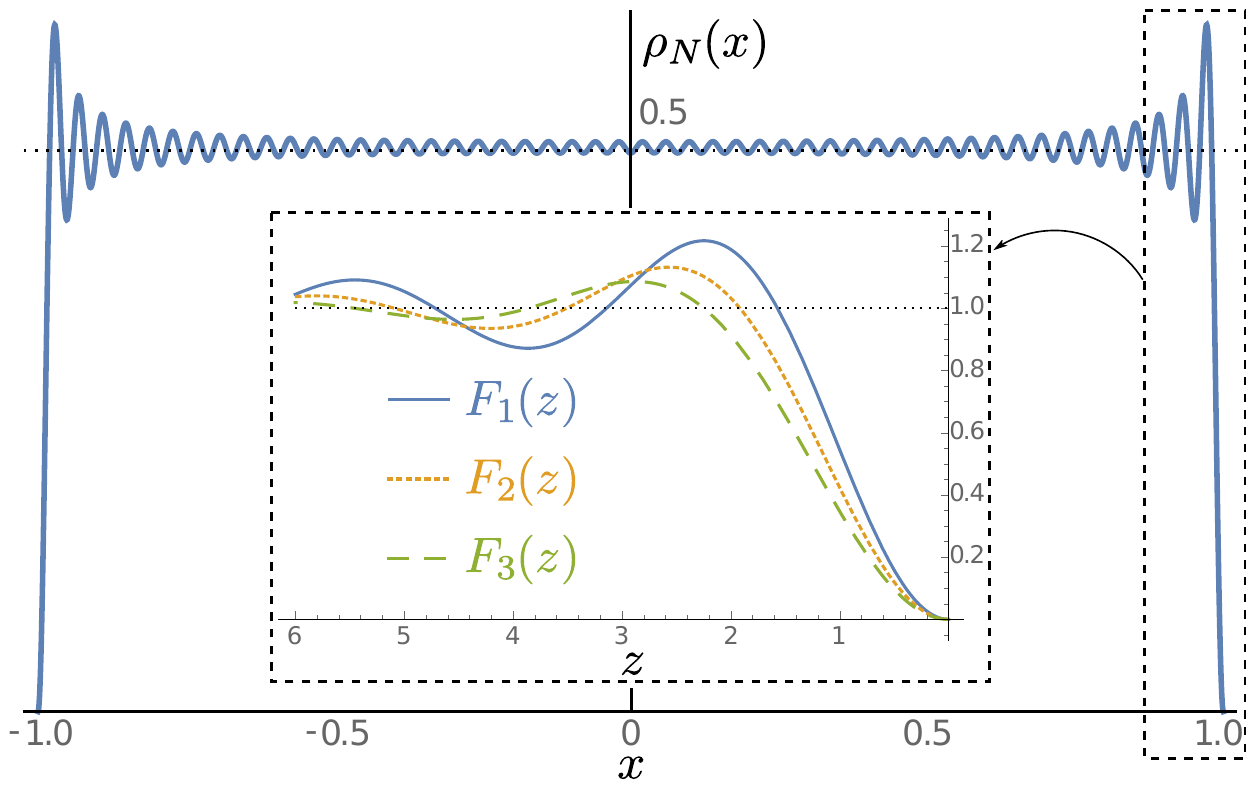}
\caption{Fermion density in a $1d$ box. Continuous lines: exact mean density for $N=50$. It vanishes near the
edge on scale $1/k_F$: the zoom (inset) indicates the scaling functions for $d=1,2,3$ as in Eq. \eqref{Fd}.}
\label{Fig_density}
\end{figure}

In this Letter, we present exact results for the edge properties of the Fermi gas in a box with hard wall potential in all dimensions and find a new universality class for the edge properties. Specifically we calculate the density, the correlations near the hard wall
as well as, in the case of a spherically symmetric potential, the distribution of the position of the fermion closest to the wall, in typical and large deviation regimes. 
We study $N$ noninteracting spinless fermions of mass $m$ in a domain ${\cal D}$, confined by a boundary $\partial {\cal D}$. We set the potential to be infinite outside ${\cal D}$. We first focus on $T=0$ and zero potential inside ${\cal D}$.
The correlations are fully characterized, thanks to the Wick theorem, by a ``kernel'' $K_\mu({\bf x},{\bf y})$, where $\mu= \q k_F^2$ is the Fermi energy. Far from the boundaries 
of ${\cal D}$, i.e. in the bulk, and for $N \gg 1$, it takes a universal, translationally invariant form \cite{fermions_review,castin,torquato}
\be \label{result_bulk}
K_{\mu}({\bf x},{\bf y}) \sim k_F^d K^{\rm b}_{d}(k_F|{\bf x}-{\bf y}|) \;,\; K^{\rm b}_{d}(r) = \frac{\J_{\frac{d}{2}}(r)}{(2\pi r)^{\frac{d}{2}}} \;,
\ee
where $\J_{\alpha}(r)$ is the Bessel function of index $\alpha$ and the superscript ${\rm b}$ refers to the bulk.
In particular, for $d=1$, $K^{\rm b}_{d=1}(r) = \sin{r}/(\pi r)$, is the sine kernel well known in RMT \cite{mehta, For10}
to describe the bulk of the spectrum. The result in \eqref{result_bulk} can be obtained using the local density approximation (LDA) \cite{castin}, or more controlled large $N$ asymptotics \cite{fermions_review,torquato}. The fermion density, given by
$\tilde \rho({\bf x})= K_{\mu}({\bf x},{\bf x})$, is thus uniform in the bulk 
$\tilde \rho({\bf x})= N \rho_N({\bf x})=\frac{N}{\Omega}=\rho_0$ with 
$\rho_0=k_F^d 2^{-d}/\gamma_d$ from \eqref{result_bulk} where $\gamma_d=\pi^{d/2} \Gamma(1+\frac{d}{2})$
and $\Omega$ the volume of the box. Hence the typical interparticle distance $\sim 1/k_F \sim (\Omega/N)^{1/d}$ is small
compared to the typical size of the box in
the limit that we study, and \eqref{result_bulk} leads to an algebraic decay of the correlations beyond that scale.

One of our main results is that near a smooth boundary point ${\bf x}_w$
the limiting kernel takes the form 
\be\label{result_edge}
K_{\mu}({\bf x},{\bf y})\sim k_F^d K^{\rm e}_{d}(k_F({\bf x} -{\bf x}_w),k_F({\bf y} - {\bf x}_w))\;,
\ee
where $K^{\rm e}_{d}$ -- the superscript `${\rm e}$' referring to the edge -- is the universal scaling function
\bea\label{edge_explicit}
K^{\rm e}_{d}({\bf a}, {\bf b}) &=& K^{\rm b}_{d}(|{\bf a} - {\bf b}|) - K^{\rm b}_{d}(|{\bf a} - {\bf b}^T|) 
\eea 
where $K^{\rm b}_{d}$ is the bulk scaling function given in \eqref{result_bulk} and ${\bf b}^T$ is the
image of ${\bf b}$ by the reflection with respect to the tangent plane to the boundary at ${\bf x}_w$. 
This is obtained by a generalized method of images, 
 which is shown to work for any smooth boundary for $N \gg 1$. This is confirmed by an exact calculation
for a spherical domain. The density near the wall is described by the scaling function 
\be 
\! \tilde \rho({\bf x}) = \rho_0 F_d(k_F z) \, , \,
F_d(z) = 1-{\small \Gamma(\frac{d+2}{2})} z^{-\frac{d}{2}}\J_{\frac{d}{2}}(2z) \label{Fd}
\ee
where $z$ is the distance of ${\bf x}$ from the boundary. It vanishes close to the wall, $F_d(z) \simeq  \frac{2}{d+2}\,z^2$ as $z \to 0$, and reaches the bulk density, $F_d(z) \to 1$, as $z \to \infty$.  This is valid for any smooth
boundary with radius of curvature $R$, in the limit $k_F R \gg 1$. Our result \eqref{result_edge}
for the kernel is thus quite different from the Airy kernel in $d=1$, and its generalizations in higher $d$,
which holds for smooth confining traps~\cite{DPMS:2015,fermions_review}. In fact in $d=1$ we show that the
positions $x_i$ of the fermions can be mapped exactly (for any $N$) to 
the eigenvalues of the Jacobi Unitary Ensemble (JUE) of RMT [see Eq. (\ref{eq: Jacobi})].
This is at variance with the corresponding exact property concerning the harmonic oscillator and the GUE~\cite{CMV2011,Eis2013,marino_prl}.

\begin{figure}[t]
\includegraphics[width=0.8\linewidth]{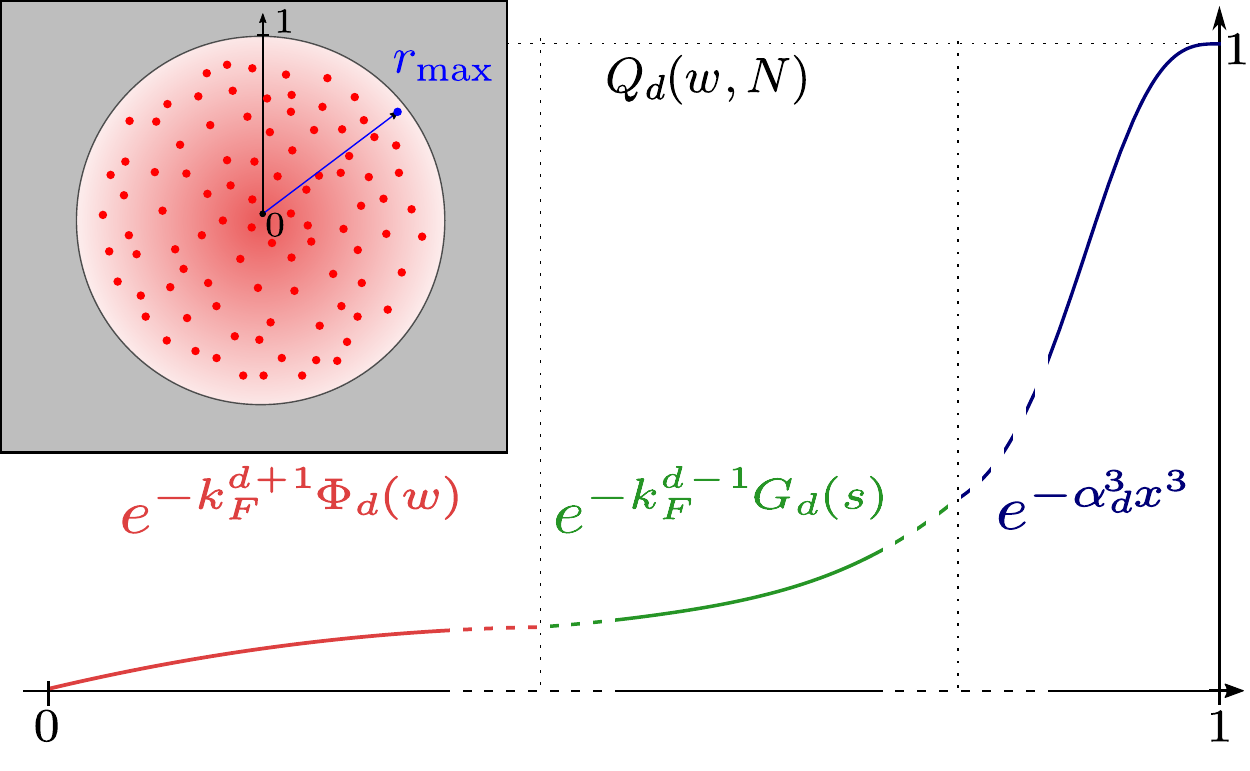}
\caption{Sketch of the cumulative distribution function $Q_d(w,N)$ of the farthest fermion position $r_{\max}$ 
in $d>1$ dimension for a spherical box as a function of $w$ and for large $N$, in the typical (blue), intermediate (green) and large deviation regimes (red), as in (\ref{summary_dgeq1}). 
\textbf{Inset}: Cartoon of a 2d Fermi gas: position of the farthest fermion indicated in blue.
}\label{Fig_sketch}
\end{figure}

As a concrete application of our result for the kernel in \eqref{result_edge}, we 
compute the cumulative distribution function (CDF) 
of the position of the farthest fermion $r_{\max} = \max_{1\leq i \leq N} \{r_i\}$ in 
a spherical box of unit radius 
in the large $N$ limit. We show that this CDF 
$Q_d(w,N) = {\rm Prob.}(r_{\max}\leq w)$, for $d>1$, displays three distinct regimes 
:  a first {\it typical} regime where $(1-w) = O(k_F^{-\frac{d+2}{3}})$, an intermediate deviation regime $(1-w) = O(k_F^{-1})$ and, a large deviation regime $(1-w) = O(1)$ (bulk). This is summarized as (see Fig. \ref{Fig_sketch})
\begin{numcases}{Q_d(w,N) \sim}
&\hspace*{-0.4cm}$e^{-[\alpha_d k_F^{\frac{d+2}{3}}(1-w)]^3} \;, \;  (1-w) = O(k_F^{-\frac{d+2}{3}})$ \nonumber\\
&\hspace*{-0.4cm}$e^{-k_F^{d-1} G_d\left(k_F\left(1-w \right) \right)}   \;,\; (1-w) = O(k_F^{-1})$ \nonumber \\
& \hspace*{-0.4cm}$e^{-k_F^{d+1} \Phi_d\left(w\right)} \;, \; (1-w) = O(1)$ \;, \label{summary_dgeq1} 
\end{numcases}
where $\alpha_d$ is a computable constant \cite{SuppMat}. The intermediate deviation function (IDF), $G_d(s)$ is
computed explicitly in \eqref{eq: G_d} and has the asymptotic behavior, $G_d(s) \sim s^3$ as $s \to 0$,
and $G_d(s) \sim s^2$ as $s \to +\infty$. The large deviation function (LDF) has the behavior
$\Phi_d(w) \sim (1-w)^2$ as $w \to 1$ and $\sim - \ln w$ as $w \to 0$. The first line of Eq.~(\ref{summary_dgeq1}), with $d>1$,
is a special case of a Weibull distribution, and is very different from the Gumbel law found 
recently for smooth potentials in $d>1$ \cite{farthest_f}. The spherical box of unit radius 
in $d>1$ reduces to the interval $[-1,1]$ in $d=1$, 
where the typical and intermediate scales coincide, $k_F^{-(d+2)/3} = k_F^{-1}$,
and the corresponding merged regime is described by the extrapolation of the
second line in Eq. (\ref{summary_dgeq1}) to $d=1$. In $d=1$, we also compute the CDF $q_1(w,N)$ of
the position $x_{\max} = \max_{1\leq i \leq N}\{x_i\}$ of the rightmost fermion. Note that $q_1(w,N) \neq Q_1(w,N)$.  
Exploiting the mapping to the JUE we show that $q_1(w,N)$
also has a typical and a large deviation regime as for $Q_{1}(w,N)$ (but no IDF) as
in \eqref{summary_dgeq1}. Most of these results generalize to a non-zero smooth potential $V({\bf x})$ inside the
box (Eqs. \eqref{result_bulk}, \eqref{result_edge}, \eqref{Fd} still hold
with $k_F \to k_F({\bf x})$ see below) and to finite temperature $T = O(\mu)$.


{\it Spherical box.} Let us start with $N$ noninteracting fermions at $T=0$, where ${\cal D}$ is 
the $d$-dimensional sphere of unit radius. The $N$ body Hamiltonian is 
${\cal H}_N = \sum_{i=1}^N H_i$ where the single particle Hamiltonian 
is defined, in spherical coordinates, as $H = -\q\Delta_{\bf x}=-\q r^{-\frac{d-1}{2}}\partial_r^2 r^{\frac{d-1}{2}}+\frac{1}{2m r^2}\hat{\bm{L}}^2$
for $r<1$, with the condition of vanishing wave-function at $r=1$. In spherical coordinates ${\bf x} = (r,{\bm \theta})$ where ${\bm \theta}$ is a $d-1$ dimensional angular vector, the eigenfunctions of $H$, using spherical symmetry, are labeled by the quantum numbers $(n,{\bf L})$, where $n$ is a positive integer, and are given by
\be
\phi_{n,{\bf L}}(r,{\bm \theta})=r^{\frac{1-d}{2}}\chi_{n,l}(r)Y_{{\bf L}}(\bm{\theta}) \,.\label{eq: decomp}
\ee  
The $Y_{{\bf L}}(\bm{\theta})$ are the $d$-dimensional spherical harmonics, labeled by the set of angular quantum numbers ${\bf L}$, which are eigenfunctions of $\hat{\bm{L}}^2$ with eigenvalues $\hbar^2 l(l+d-2)$ depending on a single positive integer $l$.
The radial part $\chi_{n,l}(r)$ is the eigenfunction of a $1d$ effective Hamiltonian, $H_{\rm eff} \, \chi_{n,l} = E_{n,l} \chi_{n,l}$, with an  effective potential 
\beq \label{Veff}
V^l_{\rm eff}(r)= \frac{\hbar^2}{2m r^2}\left(l+\frac{d-3}{2}\right)\left(l+\frac{d-1}{2}\right) \;, \; r \leq 1 
\eeq
and $V^l_{\rm eff}(r)=  +\infty$ for $r>1$. It is given by
\bea\label{chi_explicit}
\chi_{n,l}(r)= \frac{\sqrt{2r}}{\J_{\nu-1}(k_{n,l})}\J_{\nu}(k_{n,l}\,r),\ r \leq 1
\eea
where $\nu=l+\frac{d-2}{2}$ and $\chi_{n,l}(r)=0$ for $r \geq 1$. The vanishing of the wavefunction at $r=1$
determines the $k_{n,l}$'s and the eigenenergies
\bea \label{zeroes}
E_{n,l}=\q k_{n,l}^2 \quad , \quad k_{n,l} = j_{l+(d-2)/2,n}
\eea
where $j_{\nu,n}$ is the $n$-th real zero of the Bessel function $\J_\nu(x)$~\cite{dlmf}.
Each $l$-sector has degeneracy $g_d(l)$ for $d \geq 2$~\cite{harmonic_osc}
\be
g_d(0)=1,\ g_d(l)=\frac{(2l+d-2)(l+d-3)!}{l!(d-2)!},\ l>0 \;.\label{eq: degen}
\ee 

The many-body ground state wave-function $\Psi_0({\bf x}_1, \cdots, {\bf x}_N)$ is given by the Slater determinant constructed from the $N$ lowest energy single-particle wave functions (\ref{eq: decomp}) of labels $(n_1, {\bf L}_1), \cdots, (n_N, {\bf L}_N)$ as
\be\label{Slater_d}
\Psi_0({\bf x}_1, \cdots, {\bf x}_N) = \frac{1}{\sqrt{N!}} \det_{1\leq i,j \leq N} \phi_{n_i,{\bf L}_i}({\bf x}_j) \;.
\ee
For simplicity, we assume that, in the ground-state, all the levels up to the Fermi energy 
are fully occupied. We are interested in the $p$ point correlation functions
\be
R_p({\bf x}_1, ..., {\bf x}_p)=c_{N,p} \int \prod_{j=p+1}^N d{\bf x}_{j}  |\Psi_0({\bf x}_1, ..., {\bf x}_N)|^2
\ee
with $c_{N,p}= \frac{N!}{(N-p)!}$ where $|\Psi_0|^2$ is the zero temperature quantum joint PDF
of the fermion positions. Using standard manipulations \cite{fermions_review}, one can show from (\ref{Slater_d}) that all correlation functions are determinantal 
\be
R_p({\bf x}_1,\dots,{\bf x}_p)=\det_{1\leq i,j\leq p} K_\mu({\bf x}_i,{\bf x}_j) \;,\label{eq: correl_fn}
\ee
with the exact formula for the associated kernel 
\be\label{kernel_exact}
K_\mu({\bf x}, {\bf y}) = \sum_{n, {\bf L}} \phi_{n,{\bf L}}^*(r,{\bm \theta}) \phi_{n,{\bf L}}(r',{\bm \theta'}) \, \Theta(k_{F} - k_{n,l}) \;,
\ee 
where ${\Theta}(x)$ is the Heaviside step function while $k_{n,l}$ is given in (\ref{zeroes}) -- we recall that $k_F \sim N^{1/d}$. 

{\it The case of $d=1$}. In this case the fermions are confined to the segment $[-1,1]$, the eigenfunctions in Eq. (\ref{eq: decomp}) are labeled by a single quantum number $n\in\mathbb{N}^*$ and given by $\phi_n(x) = \sin({n\pi (x+1)/2)}$
%
%
%
such that $\phi_n(x=\pm 1) = 0$. Using this, the Slater determinant in (\ref{Slater_d}) 
can be evaluated explicitly and 
the quantum joint PDF of the $N$ fermions can be written as~\cite{SuppMat} (see also \cite{FFGW03,Cunden1D})
\be
|\Psi_0(x_1,..,x_N)|^2=\frac{1}{Z_N}\prod_{i=1}^N (1- v_i^2) \prod_{j <  k}^N |v_j-v_k|^2 \label{eq: Jacobi}
\ee
where $v_j = \sin \frac{\pi x_j}{2}$ and $Z_N$ is a normalization constant. Setting $u_i = (1+v_i)/2$, one finds that the $u_i$'s are distributed like the eigenvalues of random matrices belonging to the Jacobi Unitary Ensemble (JUE) \cite{mehta,For10}.
Using this connection, the scaled kernel near the hard wall at $x=1$ in Eq. (\ref{edge_explicit}) becomes \cite{CMV2011}
\be
K^{\rm e}_{1}(a,b)=\frac{\sin(a-b)}{\pi(a-b)}-\frac{\sin(a+b)}{\pi(a+b)}.\label{eq: K_1d}
\ee
Using this kernel we evaluate the CDF, $q_1(w,N)$, of the position $x_{\max}$ of the rightmost fermion,
by observing that this is the "hole probability" that there are no fermions in the interval $[w,1]$. In RMT, it is
well known that such hole probabilities can be formally expressed as a Fredholm determinant, with an associated kernel.
In our case, we can then express $q_1(w,N)$ as a Fredholm determinant with the kernel given in \eqref{eq: K_1d}, $q_1(w,N) = {\rm Prob.}(x_{\max} \leq w) \sim \tilde q_1(k_F(1-w))$ with
\be\label{eq: P_1d}
\hspace*{-0.2cm}\tilde q_1(s)= D_-\left(\frac{2s}{\pi}\right) \,, \, D_-(t) = \Det\left(I-P_{\frac{\pi\,t}{2}} K^{\rm e}_{1}P_{\frac{\pi t}{2}}\right) 
\ee
where $P_x$ denotes the projector on the interval $[0,x]$. Interestingly, 
$D_-(t)$ also describes the probability to find at most one eigenvalue in the interval $[0,t]$ (in unit of the average spacing) in the bulk of the spectrum of matrices belonging to the Gaussian Orthogonal Ensemble (GOE)~\cite{mehta,DysonD,TracyD}. 
%
%
Finally, for large deviations, $1-w \gg k_F^{-1}$ (where $k_F \sim N$), we show \cite{SuppMat}, using Eq. (\ref{eq: Jacobi}), 
that $q_1(w,N) \sim \exp(- k_F^2 \varphi_1(w))$ with
\be
\varphi_1(w)=-\frac{4}{\pi^2}\ln\left(\frac{1}{2}+\frac{1}{2}\sin\left(\frac{\pi}{2}w\right)\right) \;,Ê\; -1 \leq w \leq 1 \;. \label{eq: large_dev_1d}
\ee

{\it The case $d>1$.} In this case, although there is no obvious connection with RMT, the positions of the fermions form a $d$-dimensional determinantal process with a kernel obtained by substituting Eqs. (\ref{eq: decomp}) and (\ref{chi_explicit}) in 
(\ref{kernel_exact}).
In the large $N$ limit, using known
asymptotics of the Bessel functions and their zeroes,  we derive~\cite{SuppMat} the limiting form of the kernel in Eq. (\ref{edge_explicit}). 
%
%

Using the determinantal form of $\Psi_0$ in (\ref{Slater_d}) and the spherical symmetry of the problem,
following similar steps as in \cite{farthest_f} we show that the CDF of $r_{\max}$ factorizes
as a product over different angular sectors 
\be\label{Qd_explicit_2}
Q_d(w,N) = \prod_{l=0}^{l_N^*} \left[ P_{l}(w,m_l)\right]^{g_{d}(l)} \;,
\ee
where $g_d(l)$ is the degeneracy of each $l$-sector (\ref{eq: degen}) and $m_l= \sum_n \Theta(k_F-k_{n,l})$ is the number of fermions in this sector. In Eq. (\ref{Qd_explicit_2}), $l_N^*$ is the last occupied $l$-sector, with 
$l_N^* \to k_F$ in the large $N$ limit \cite{SuppMat}. In Eq.~(\ref{Qd_explicit_2}), $P_{l}(w,m_l)$ is the CDF of the position $x_{\max,l}$ of the rightmost fermion among $m_l$ fermions in the $1d$ effective potential $V^l_{\rm eff}(r)$ in Eq. (\ref{Veff}). 
Eq.~(\ref{Qd_explicit_2}) shows that $r_{\max}$ is the maximum among a large number of independent but non identical random variables $x_{\max,l}$, each counted with its degeneracy $g_d(l)$. In the large $N$ limit, the product in Eq. (\ref{Qd_explicit_2}) is dominated by large values of $l = O(k_F) = O(N^{1/d})$, corresponding to large values
of $m_l \sim k_F $ \cite{SuppMat}. Because of the hard-wall potential at $r=1$, $x_{\max,l}$ is bounded by 1.
Close to the wall, one finds that the CDF $P_l(w,m_l)$ of $x_{\max,l}$ behaves for large $l = \tilde l k_F$, with $\tilde l$ fixed, as $P_{l}(w,m_l) \sim \tilde q_1(k_F(1-w)\sqrt{1-\tilde l^2})$. Using $\tilde q_1(s) \sim 1 - 2\,s^3/(9\pi)$ as $s \to 0$ \cite{SuppMat}, 
we find that the PDF of $x_{\max,l}$, $\partial_w P_l(w,m_l) \propto (1-w)^2$ when $w \to 1$. Had these variables been 
identically distributed with this PDF, then from the classical theory of EVS, their maximum $r_{\max}$ would be distributed by the Weibull law $Q_d(w,N) \sim \exp[- a_N(1-w)^3]$ for some $a_N$. The exact result in the first line of (\ref{summary_dgeq1}) thus demonstrates that effectively these variables become independent.

 As in $d=1$, there is a large deviation regime for $(1-w) = O(1)$, i.e. $Q_d(w,N) \sim \exp[-k_F^{d+1} \Phi_d(w)]$ [see the third line of Eq. (\ref{summary_dgeq1})]. While computing $\Phi_d(w)$ remains a challenge, its small $w$ behavior can be obtained as $\Phi_d(w) \sim_{w \to 0} - \kappa_d \ln w$~\cite{SuppMat}.
 Hence for $d>1$, this large deviation regime can not match with the typical fluctuations where $Q_d(w,N) \sim e^{- k_F^{d+2} (\alpha_d(1-w))^3}$, for $1-w = O(k_F^{-\frac{d+2}{3}})$. Indeed, there is a new intermediate regime, for $(1-w) = O(k_F^{-1})$,
 which can be obtained from the exact formula (\ref{Qd_explicit_2}). It is reminiscent of the typical fluctuations of $x_{\max,l}$ within each $l$-sector, and one finds that $Q_d(w,N) \sim e^{-k_F^{d-1} G_d(k_F(1-w))}$ where 
\be
G_d(s)=-\int_{0}^{1}\, \frac{2\, \tilde l^{d-2} }{\Gamma(d-1)}\ln D_{-}\left(\frac{2}{\pi}s\sqrt{1-\tilde l^2}\right)d\tilde l \;, \label{eq: G_d}
\ee
with $D_-(t)$ given in Eq. \eqref{eq: P_1d}. Using the asymptotic properties of $D_-(t)$ (see \cite{SuppMat}) we find that $G_d(s) \sim s^3$ as $s \to 0$,
and $G_d(s) \sim s^2$ as $s \to +\infty$. This ensures a smooth matching between the three regimes in~\eqref{summary_dgeq1}.

{\it General domain.} We now consider a general domain ${\cal D}$ with the single particle Hamiltonian $H= -\q \Delta_{{\bf x}} + V({\bf x})$ 
with $V({\bf x}) = 0$ if ${\bf x} \in {\cal D}$ and $V({\bf x}) = +\infty$ outside ${\cal D}$. To derive~(\ref{edge_explicit}) we use the representation~\cite{fermions_review}
\be \label{fromGtoK2} 
K_\mu({\bf x}, {\bf y}) = \int_{{\cal C}} \frac{dt}{2 i \pi t} \exp\left(\frac{\mu t}{\hbar}\right) G({\bf x}, {\bf y};t)
\ee 
where ${\cal C}$ is the Bromwich contour in the complex plane and $G({\bf x}, {\bf y};t)$ 
is the euclidean propagator associated to $H$. Here it is the solution of
the free diffusion equation $- \hbar \partial_t G = -\q \Delta_{{\bf y}} G$
inside ${\cal D}$ with Dirichlet condition $G=0$ on $\partial{\cal D}$.
Let us start with a box $]-R,R[$ in $d=1$. Using the standard method of images
to express $G$ and using \eqref{fromGtoK2} leads to the exact result \cite{SuppMat}
\be \label{Kimages} 
K_\mu(x,y)=  k_F \sum_{\substack{\epsilon=\pm 1\;, n \in \mathbb{Z}}}   \epsilon K^{\rm b}_1(k_F |x- \epsilon y - (4 n+1-\epsilon) R|) 
\ee
where $\mu=\frac{\hbar^2}{2 m} k_F^2$ and $K^b_1$ is the sine kernel (\ref{result_bulk}). Because 
$K_1^{\rm b}$ decays, in the limit $k_F R \gg 1$ at most one image 
contributes and one recovers (\ref{eq: K_1d}). This is a general feature, and
for large $\mu$ in any $d$ e.g. with planar walls, only images within $1/k_F$ 
need to be considered.


Consider for simplicity the case of fermions in a disc ${\cal D}$ of radius $R$ in $2d$ with Dirichlet boundary conditions on the circular boundary $\partial{\cal D}$, parameterized by $(y_1,y_2)$ measured with respect to an origin $(0,0)$ chosen on $\partial {\cal D}$. The circular boundary is described by $y_2^2 + (R-y_1)^2 = R^2$. Locally,
near the origin $(0,0)$, where $y_1, y_2 \ll 1$, one has $y_1 \simeq y_2^2/2R$. The key idea then is to use the fact 
that for large $\mu$
the time scale which dominates the integral 
in \eqref{fromGtoK2} is $t \sim t^*=\hbar/\mu \ll 1$. 
It is then natural to use the rescaled time $\tilde t= \mu \, t/\hbar$ and correspondingly rescaled space 
$\tilde {\bf x}= k_F {\bf x}$ and to rewrite \eqref{fromGtoK2} in terms of the rescaled
propagator $\tilde G$, such that $G({\bf x}, {\bf y}, t) = k_F^d \tilde G(\tilde {\bf x}, \tilde {\bf y},\tilde t) $. The latter must now vanish on the curve $\tilde y_1 \simeq \tilde y_2^2/(2 k_F R)$
(see \cite{SuppMat} for details). Hence in the limit $k_F R \gg 1$ the wall can be effectively replaced by a straight line
and the method of images applies for $\tilde {\bf x} \sim \tilde {\bf y} = O(1)$, i.e.
within a distance $1/k_F$ from the wall, leading to the general formula~\eqref{edge_explicit}.

We note that the argument using the method of images
can be extended in several directions. This includes fermions in any $d$-dimensional 
domain ${\cal D}$ (with zero potential inside) with a smooth boundary $\partial {\cal D}$, provided $k_F R \gg 1$ where $R$ 
is the minimum local radius of curvature of $\partial {\cal D}$. It also holds, under certain conditions,  
in the case when the potential
inside ${\cal D}$ is nonzero~\cite{SuppMat}. Finally, this
method of images can also be generalized to finite temperature $T \sim \mu$, using the 
finite temperature bulk kernel derived in \cite{fermions_review} (see \cite{SuppMat}).

The above argument fails, however, for a wedge or a cone with apex point $O$
at which the radius of curvature is ill-defined. Consider e.g. a 
wedge domain in $d=2$ with angle $\alpha$. 
For $\alpha=\frac{2 \pi}{m}$ and integer $m$, the method of images can again be used
\cite{SuppMat}. A more general formula exists for any $\alpha$ for the propagator $G$ \cite{cone,wedge1},
leading to an exact formula for the kernel \cite{SuppMat}, using \eqref{fromGtoK2}. 
Let us display only the small distance behavior near $O$ in polar coordinates ${\bf x}=(r,\phi)$, ${\bf x}_0=(r_0,\phi_0)$
\be \label{wedgeKsmall} 
K_\mu({\bf x}, {\bf x}_0) \sim c_\alpha^{-1} \sin \left(\frac{\pi \phi}{\alpha}\right) 
\sin\left(\frac{\pi \phi_0}{\alpha}\right) k_F^2 (k_F r_0 r)^{\frac{\pi}{\alpha} }
\ee
with $c_\alpha=\pi 2^\frac{2\pi}{\alpha} \Gamma(\frac{\pi}{\alpha} ) \Gamma(2+  \frac{\pi}{\alpha})$,
hence the density vanishes as $\sim r^{2 \pi/\alpha}$ near $O$, but quadratically along
the walls. Similar expressions exist for any conical box in any $d$~\cite{SuppMat}.

In conclusion, we have computed exactly the $d$ dimensional kernel that characterizes the
correlation functions at the edge of a Fermi gas close to a hard wall in $d \geq 1$ dimensions.
The density vanishes algebraically near the wall, with an exponent $2$ (quadratically) near 
a smooth wall, while near a cone the exponent depends continuously on the solid angle of the cone. We have also 
obtained the exact distribution of the position of the fermion closest to the wall
and found three regimes in $d \geq 1$, while only two regimes for $d=1$. Given the 
ever increasing sophistication of designing atomic traps of various shapes, it would
be interesting to test experimentally our exact theoretical predictions for the edge correlations in a hard box
potential.

%
%
%
%
%
%
%
%
%
%

{\it Acknowledgments:} We thank D. S. Dean and J. Grela for useful discussions.

{}

\newpage


\newpage

\begin{widetext} 

\bigskip

\bigskip

\begin{large}
\begin{center}

Supplementary Material for {\it Statistics of fermions in a $d$-dimensional box near a hard wall}

\end{center}
\end{large}

%
%
%
%
%
%
%
%


\section{A) $d-$dimensional spherically symmetric hard box potential: exact results}

Our starting point is a simple model of $N$ non-interacting fermions in a $d$-dimensional spherically symmetric hard box potential.
The Hamiltonian of the system is as described in the main text 
\bea\label{eq:HN_supp}
{\cal H}_N = \sum_{i=1}^N H_i \;, \; H_i = -\q \Delta_{{\bf x}_i} + V_W({\bf x}_i) \;,
\eea
where $V_W({\bf x}) = V_W(|{\bf x}|)$ is a spherically symmetric hard box potential given by
\bea\label{def_VW_supp}
V_W(|{\bf x}|) = 
\begin{cases}
&0 \;, \; {\bf x} \in {\cal{D}}=\lbrace{\bf x}\in \mathbb{R}^d,\abs{\bf{x}}\leq 1\rbrace \\
&+\infty \;, \; {\bf x} \notin {\cal{D}} \;.
\end{cases}
\eea
Many exact results can be derived in this model, as used in the main text, and details are provided here.

\subsection{A. 1) Exact solution for $N$ fermions in a box in $d=1$}

\subsubsection{Mapping to the Jacobi Unitary Ensemble}

In $1d$, the hard wall potential $V_W(x)$ in Eq. (\ref{def_VW_supp}) confines the fermions on the interval $[-1,1]$. In this case, the single particle eigenfunctions $\phi_n(x)$, with boundary conditions $\phi_n(x=\pm 1) = 0$, are given by
\be\label{phin_supp}
\phi_n(x)=\sin\left(\frac{n\pi}{2}(x+1)\right),\ -1\leq x\leq 1 \;,
\ee
when $n>0$ is a positive integer. The associated eigenenergies $E_n$ read
\be
E_n=\q k_n^2 \;, \;{\rm with}\;\; k_{n}=\frac{n\pi}{2} \;.
\ee
In the $N$-body ground state of the fermions, each single particle level $n=1$ up to $n=N$ are filled by a single fermion. 
The highest occupied level (Fermi level) has energy $\mu = E_N = \hbar^2 k_F^2/(2m)$ with $k_F=\frac{N\pi}{2}$.

The ground state wave function $\Psi_0(x_1, \cdots, x_N)$ is given by the Slater determinant constructed from the $N$ lowest single particle eigenstates (\ref{phin_supp}) as 
\be\label{GS_1d_supp}
\Psi_0(x_1, \cdots, x_N) = \frac{1}{\sqrt{N!}} \det_{1\leq i,j \leq N} \phi_j(x_i) \;.
\ee
One can rewrite this Slater determinant using the following simple identity 
\be\label{identity_U}
\sin n \theta =\sin \theta\ U_{n-1}(\cos \theta),\quad {\rm with}\quad U_{n}(t)=\sum_{p=0}^{\lfloor\frac{n}{2}\rfloor}\dbinom{n+1}{2p+1}t^{n-2p}\left(1-t^2\right)^{p} \;,
\ee 
where $U_n(t)$ are called  the Chebyshev polynomials of second kind \cite{dlmf_supp}. In Eq. (\ref{identity_U}) 
$\lfloor n/2 \rfloor$ denotes the greatest integer less than or equal to $n/2$. We now use this identity (\ref{identity_U}) in Eq. (\ref{GS_1d_supp}) with $\theta=\frac{\pi}{2}(x+1)$, such that $\cos(\theta) = -\sin(\pi x/2)$. The rows and columns of the 
Slater determinant in (\ref{GS_1d_supp}) can be rearranged to express this determinant in terms of a Vandermonde determinant. This then yields the following expression for the quantum probability distribution function (PDF)
\be\label{qPDF_1d_supp}
\abs{\Psi_0(x_1, \cdots, x_N)}^2=\frac{1}{Z_N}\prod_{i=1}^N \cos^2\left(\frac{\pi x_i}{2}\right)\prod_{j\leq k}^N \abs{\sin\left(\frac{\pi x_j}{2}\right)-\sin\left(\frac{\pi x_k}{2}\right)}^2.
\ee
Here, $Z_N$ is a normalization constant and can be computed exactly~\cite{Mehta_supp}
\be\label{ZN_supp}
Z_N=2^{N^2}\left(\frac{4}{\pi}\right)^N Z_N',\ {\rm with}\ Z_N'=\prod_{j=1}^N\frac{\Gamma(1+j)\Gamma\left(\frac{1}{2}+j\right)^2}{\Gamma(N+1+j)}\ .
\ee
By performing the change of variables $u_i=\frac{1}{2}(1+\sin\left(\frac{\pi x_i}{2}\right))$, one can bring this quantum joint PDF to the standard form of the joint distribution of the Jacobi Unitary Ensemble~\cite{Mehta_supp}	
\be\label{JUE}
{\rm Prob}.(u_1,\cdots,u_N)=\frac{1}{Z_N'}\prod_{i=1}^N \sqrt{u_i(1-u_i)}\prod_{j\leq k}^N \abs{u_j-u_k}^2 \;, \;\; {\rm where} \;\; 0 \leq u_i \leq 1 \;.
\ee

The $p$-point correlation function $R_p({x}_1, \cdots, {x}_p)$ is defined as
\be
R_p({x}_1, \cdots, {x}_p)= \frac{N!}{(N-p)!} \int |\Psi_0({x}_1,\cdots,x_{p}, x_{p+1}, \;, \cdots{x}_N)|^2 \; dx_{p+1} \cdots dx_N \;.
\ee
Using standard manipulations of RMT \cite{Mehta_supp,fermions_review_supp}, one can show that correlation functions are determinantal for all $p= 1, \cdots, N$
\be
R_p({x}_1,\dots,{x}_p)=\det_{1\leq i,j\leq p} K_\mu({x}_i,{x}_j) \;,\label{eq: correl_fn_supp}
\ee
where the kernel $K_\mu({x},{y})$ is given by  
\be\label{kernel_exact_1d_supp}
K_\mu({x}, {y}) = \sum_{n=1}^N \phi_{n}^*(x) \phi_{n}(y)  = \frac{\sin\left(\frac{(2N+1)\pi}{4}(x-y)\right)}{2\sin\left(\frac{\pi}{4}(x-y)\right)}-\frac{\sin\left(\frac{(2N+1)\pi}{4}(2+x+y)\right)}{2\sin\left(\frac{\pi}{4}(2+x+y)\right)} \;,
\ee 
where we have used $\phi_n(x)$ from Eq. (\ref{phin_supp}).

We now consider the scaling behavior of this kernel $K_\mu({x}, {y})$ at the edge near $x=1$ (or near its symmetric counterpart at $x=-1$), for large $\mu \sim N^2$. For this, we set the distances $\abs{x-1}\sim k_F^{-1}$ and $\abs{y-1}\sim k_F^{-1}$, with $k_F = N\pi/2$, denoting the inverse of the typical inter-particle distance at the edge. Substituting $x-1 = a/k_F$ and $y-1 = b/k_F$ and expanding for large $k_F \sim N$, one obtains the leading order scaling behavior of the kernel 
\be\label{1d_k_supp}
K_{\mu}(x,y)=k_F K_1^{\rm e}(k_F(x-1),k_F(y-1)),\ {\rm with}\ K_1^{\rm e}(a,b)=\frac{\sin(a-b)}{\pi(a-b)}-\frac{\sin(a+b)}{\pi(a+b)}\;,
\ee
as mentioned in Eq. (16) of the main text. 

\subsubsection{Rightmost fermion CDF $q_1(w,N)$}

{\it Typical fluctuations of $x_{\max}$}. Consider the CDF of the position of the rightmost fermion $q_1(w,N) = {\rm Prob.}(x_{\max} \leq w)$. The event that $x_{\max} \leq w$ necessarily indicates that there are no fermions in the interval $[w,1]$. Thus this can be interpreted as a ``hole probability'' (i.e., an interval free of particles). In RMT, this hole probability is a standard observable and it is well known that it can be expressed as a Fredholm determinant with an associated kernel (it can also be expressed in terms of the solution of a Painlev\'e VI equation \cite{TW_painleve,Selberg_jacobi_supp}). In our case, in the limit of large $N$, and for $w-1 \sim k_F^{-1}$, $q_1(w,N)$ has a scaling form 
\be\label{scaling_q_supp}
q_1(w,N)\to \tilde{q}_1(k_F(1-w))\;,\quad {\rm with}\quad \tilde{q}_1(s)=\Det\left(I-P_{s}K_1^{\rm e}P_{s}\right)\;,
\ee
where $\tilde{q}_1(s)$ is a Fredholm determinant with kernel $K_1^{\rm e}$ given in Eq. (\ref{1d_k_supp}). The notation 
$P_s$ denotes the projector on $\left[ 0,s\right]$. As stated in the main text, one can also write that $\tilde q_1(s) = D_-(2\,s/\pi)$ where $D_-(t)$ is a Fredholm determinant that appears in the classical Gaussian Orthogonal Ensemble in a different context \cite{Mehta_supp}.

{\it Atypical fluctuations of $x_{\max}$.} As we have seen in the previous paragraph, typical fluctuations of $x_{\max}$ near $w=1$ occur on a scale $1-w \sim {k_{F}}^{-1}$ (where $k_F \sim N$ is large) and the CDF of such typical fluctuations are described by the scaling form in Eq. (\ref{scaling_q_supp}). What about the fluctuations that are atypically large, e.g. when $1-w=O(1)$. The scaling form in (\ref{scaling_q_supp}) can not be applied in this regime and we need a separate calculation. 
This can be done very simply as follows. We start from the joint distribution of the $u_i = \frac{1}{2}\left(1+\sin(\pi \,x_i/2) \right)$ variables in Eq. (\ref{JUE}). The event $x_{\max} \leq w$ corresponds to $u_{\max} \leq g(w) = \frac{1}{2}(1+\sin\left(\frac{\pi w}{2}\right))$. Hence we can write
\be\label{large_dev_comp}
\hspace*{-1.3cm} q_1(w,N)=\frac{1}{Z_N'}\int_{0}^{g(w)}du_1\cdots\int_{0}^{g(w)}du_N\prod_{i=1}^N \sqrt{u_i(1-u_i)}\prod_{j\leq k}^N \abs{u_j-u_k}^2\;, \quad {\rm where} \quad g(w) = \frac{1}{2}(1+\sin\left(\frac{\pi w}{2}\right)) \;,
\ee
and $Z_N'$ given in Eq. (\ref{ZN_supp}). In the limit of large $N$, the dominant contribution to this multiple integral comes from the squared Vandermonde term (and is of order $e^{N^2}$), while the product $\prod_{i=1}^N \sqrt{u_i(1-u_i)}$ is of order $e^{N}$. Hence, keeping only the squared Vandermonde term, and neglecting the rest, we can estimate the remaining integral simply by a change of variables $v_i = u_i/g(w)$ followed by a power counting. This gives to leading order for large $N$
\be\label{large_dev_comp2}
 q_1(w,N) \sim \left[ g(w)\right]^{\frac{N^2}{2}} \;,
\ee
which can be re-written in the large deviation form (using $k_F = N\pi/2$)
\be\label{large_dev_1d_supp}
q_1(w,N)\sim \exp\left(-k_F^2\varphi_1(w)\right)\;,\quad {\rm with} \quad \varphi_1(w)=-\frac{4}{\pi^2}\ln\left(\frac{1}{2}+\frac{1}{2}\sin\left(\frac{\pi}{2}w\right)\right)\;,
\ee
as announced in Eq. (29) of the main text. The rate function $\varphi_1(w)$ is plotted in Fig. \ref{Fig_phi} and it has the following asymptotic behaviors 
\bea
\varphi_1(w)\to
\begin{cases}
&\frac{(w-1)^2}{4},\; \quad \quad \quad \quad \; w\to 1\;,\\
&-\frac{8}{\pi^2}\ln(1+w),\; \quad w\to -1\;.
\end{cases}
\eea
Using this asymptotic behavior of $\varphi_1(w)$ when $w \to 1$, one gets $q_1(w,N) \sim e^{-k_F^2 \frac{(w-1)^2}{4}}$. In contrast, if we start from the typical regime in Eq. (\ref{scaling_q_supp}), using the asymptotic behavior of $\tilde q_1(s)$ as $s \to \infty$ (i.e., $\tilde q_1(s) \sim e^{-s^2/4}$), we find $q_1(w,N)  \sim e^{-k_F^2 \frac{(w-1)^2}{4}}$ to leading order. Thus we see that there is a smooth matching between the typical regime and the large deviation regime.

\begin{figure}[h]
\includegraphics[width=0.45\textwidth]{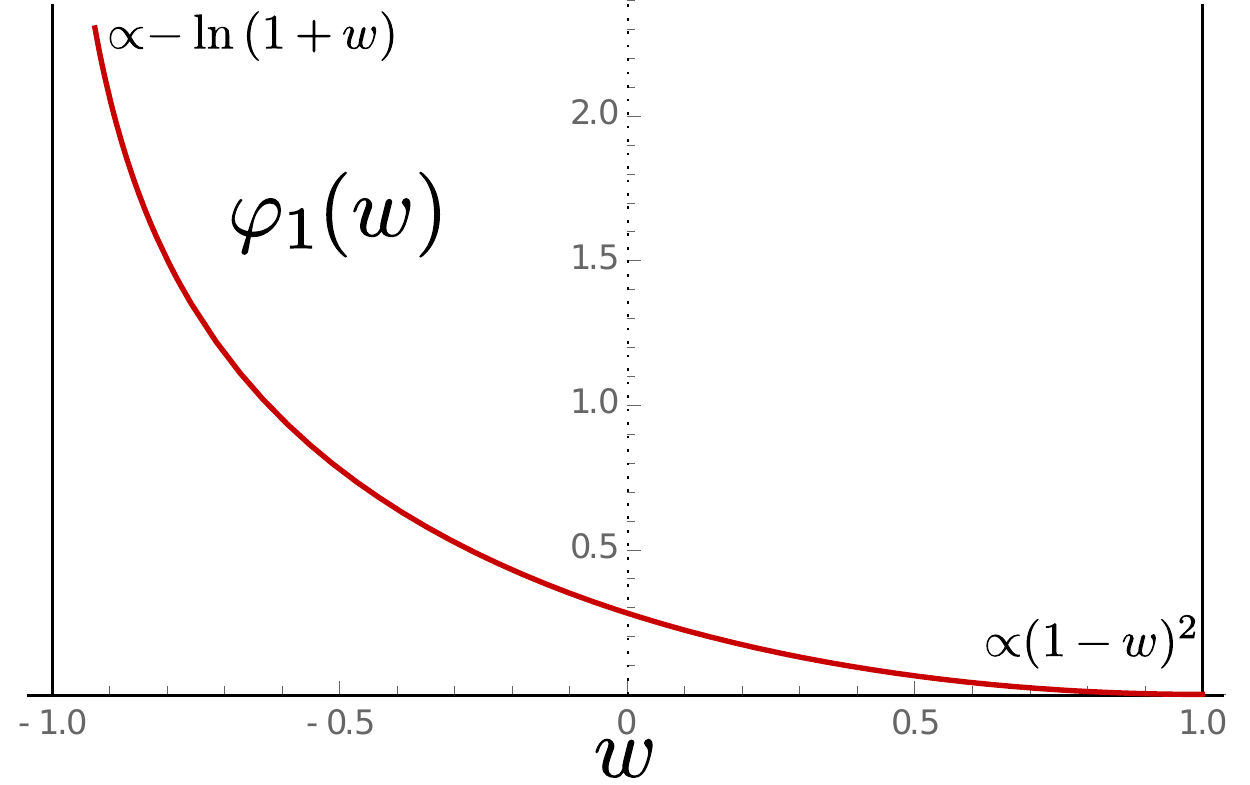}
\caption{Large deviation function $\varphi_1(w)$ given in Eq. (\ref{large_dev_1d_supp}) corresponding to atypically large fluctuations of $x_{\max}$
in $d=1$.}\label{Fig_phi}
\end{figure}

\subsection{A. 2) Exact solution for $N$ fermions in a spherical box in $d>1$}

\subsubsection{Computation of the finite $N$ kernel}

Our starting point is the exact expression for the $d$-dimensional kernel $K_{\mu}({\bf x},{\bf y})$ in spherical coordinates ${\bf x}=(r,{\bm \theta})$, ${\bf y}= (r',{\bm \theta'})$ given in the main text in Eq. (14) together with Eqs. (6) and (8)
\bea\label{Kmu_supp1}
K_{\mu}({\bf x},{\bf y}) =  (r\,r')^{\frac{1-d}{2}} \sum_{n,{\bf L}} Y_{\bf L}^*({\bm \theta})Y_{\bf L}({\bm \theta'}) {\chi}^*_{n,l}(r) {\chi}_{n,l}(r')  \Theta(k_F-k_{n,l}) \;,
\eea
where $\Theta(x)$ is the Heaviside step function, $k_F = \sqrt{2\,m\,\mu/\hbar^2}$ with $\mu$ being the Fermi energy and $k_{n,l} = j_{l+(d-2)/2,n}$ where $j_{\nu,n}$ is the $n$-th real zero of the Bessel function ${\rm J}_\nu(x)$. The function $\chi_{n,l}(r)$ in Eq. (\ref{Kmu_supp1}) is given by
\bea\label{chi_explicit_supp}
\chi_{n,l}(r)= \frac{\sqrt{2r}}{\J_{\nu-1}(k_{n,l})}\J_{\nu}(k_{n,l}\,r),\ r \leq 1
\eea
where $\nu=l+\frac{d-2}{2}$ and $\chi_{n,l}(r)=0$ for $r \geq 1$. To analyze the discrete sums in Eq. (\ref{Kmu_supp1}), it is convenient to parameterize the set of quantum numbers ${\bf L}$ as ${\bf L}=(l,\bm{m})$ where $\bm{m}$ is a $d-2$ dimensional vector which, for a given value of $l$, takes $g_d(l)$ different values corresponding to distinct eigenstates [see Eq. (\ref{eq: degen}) of the main text] which have all the same eigenenergy $E_{n,l} = \hbar^2 k_{n,l}^2/(2m)$. 
Because of the step function $\Theta(k_F-k_{n,l})$, both $l$ and $n$ are bounded (see Fig. \ref{Fig4_supp}): within each $l$-sector, $n$ is bounded by $m_l$, $0\leq n \leq m_l$ (see Fig. \ref{Fig4_supp}), where $m_l$ is given by  
\be\label{ml_supp}
m_l = \sum_{n\geq 1} \Theta(k_F-k_{n,l}) \;, \quad  {\rm or \; equivalently} \;\quad  k_{m_l,l} \leq k_F <  k_{m_l+1,l} \;.
\ee 
Similarly $l$ is bounded by $l_{N}^*$, $0\leq l \leq l_N^*$ (see Fig. \ref{Fig4_supp}) such that $m_{l_N^*} > 0$ and $m_{l_N^* + 1} = 0$. Here, for simplicity, we restrict ourselves to non-degenerate ground state, i.e., the highest energy level is fully occupied and therefore the total number of fermions $N$ is given by (using the parameterization ${\bf L}=(l,\bm{m})$)
\be\label{N_supp1}
N =  \sum_{n,{\bf L}}  \Theta(k_F-k_{n,l}) = \sum_{l \geq 0} \sum_{\bm m} \sum_{n \geq 1} \Theta(k_F-k_{n,l}) = \sum_{l=0}^{l_N^*} g_d(l) \, m_l \;.
\ee 
Using this parameterization of the quantum numbers ${\bf L} = (l,{\bf m})$, the kernel in Eq. (\ref{Kmu_supp1}) reads
\be\label{d_kernel}
K_{\mu}({\bf x},{\bf y})=(r\,r')^{\frac{1-d}{2}}\sum_{l=0}^{l_N^*}\left(\sum_{\bm{m}}Y_{l,\bm{m}}^*({\bm \theta})Y_{l,\bm{m}}({\bm \theta'})\right)K^l_{\rm eff}(r,r') \;,
\ee
where $K^{l}_{\rm eff}(r,r')$ is the effective one-dimensional kernel for fermions in a given $l$-sector
\be
K^{l}_{\rm eff}(r,r')=\sum_{n=1}^{m_l}\chi_{n,l}^*(r)\chi_{n,l}(r') \;,\label{eq: K_eff_1d_supp}
\ee
and is completely independent of $\bm{m}$. 
%
%
%
%
%
\begin{figure}[ht]
\includegraphics[width=0.5\linewidth]{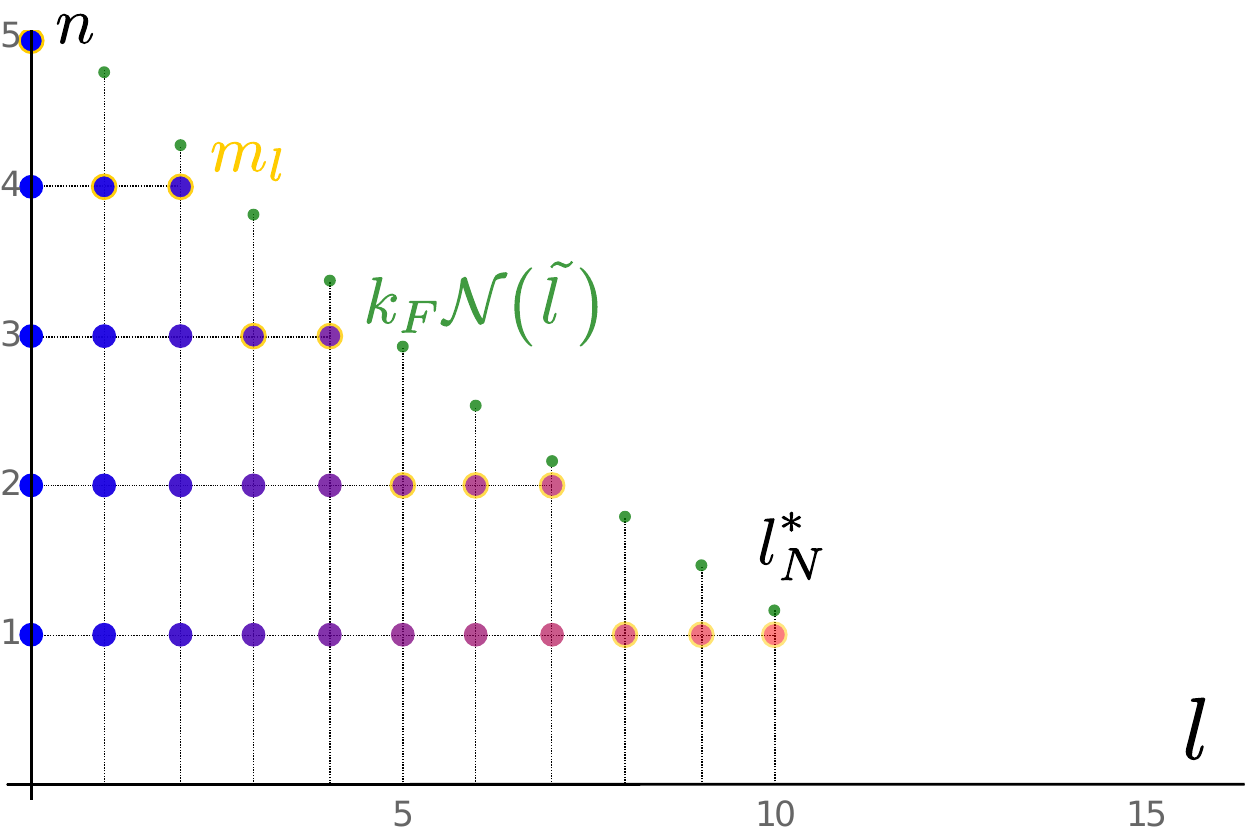}
\caption{Occupation of the energy levels in the $(l,n)$ plane in the ground-state of the $2d$ spherical hard box for $k_F = 15$, which corresponds to $N = 56$ fermions (in good agreement with the asymptotic formula $N=(k_F/2)^2\approx 56$). In this case, the energy levels are $E_{n,l} = \hbar^2 j_{l,n}^2/(2m)$ where $j_{\nu,n}$ is the $n$-th real zero of the Bessel function ${\rm J}_\nu(x)$ and the degeneracy is $g_{d=2}(l) = 2$ for all $l$. The filled circles indicate the occupied states (note that each circle actually corresponds to two distinct quantum states, as $g_{n,l} =2$). In each $l$-sector, each one indicated with a different color, there are $m_l$ occupied states and the last occupied state is indicated in yellow. The green points correspond to the asymptotic large $N$ behavior $m_l \sim k_F{\cal N}(\tilde l = l/k_F)$ as given in Eq. (\ref{m_l_supp}) and they are in relatively good agreement with the exact values of $m_l$, even for this low value of $k_F$. Finally, $l_N^* = 10$ denotes the last occupied $l$-sector, i.e. $m_l = 0$ for $l > l_N^*$.}\label{Fig4_supp}
\end{figure}
In Eq. (\ref{d_kernel}), the sum over the quantum numbers ${\bm m}$ runs over $g_d(l)$ possible values (as the ground-state is non-degenerate) and it can be performed explicitly using the following sum rule for the spherical harmonics~(see for instance~\cite{harmonics}) 
\be\label{sum_harm_supp}
\sum_{\bm{m}}Y_{l,\bm{m}}^*({\bm \theta})Y_{l,\bm{m}}({\bm \theta'})=\frac{g_d(l)}{S_d}{\sf P}_{l,d}(t) \;, \; t = \frac{{\bf x}\cdot {\bf y}}{|{\bf x}|\,|{\bf y}|} \;,
\ee
where $S_d = d \, \pi^{d/2}/\Gamma(d/2+1)$ is the area of the $d$-dimensional unit sphere and where the variable $t$ depends obviously only on the angular variables ${\bm \theta}$ and ${\bm \theta'}$. The functions ${\sf P}_{l,d}(t)$ are related to the associated Legendre polynomials \cite{harmonics} and satisfy the differential equation
\be
(1-t^2){\sf P}_{l,d}''(t)+(1-d)t\,{\sf P}_{l,d}'(t)+l(l+d-2){\sf P}_{l,d}(t)=0 \;, \label{eq:Legendre} 
\ee
with the conditions ${\sf P}_{l,d}(-t)=(-1)^l {\sf P}_{l,d}(t)$ and ${\sf P}_{l,d}(1)=1$. 
Finally, the kernel takes the simplified form
\be\label{kernel_finite_supp}
K_{\mu}({\bf x},{\bf y})=\frac{(r\,r')^{\frac{1-d}{2}}}{S_d}\sum_{l=0}^{l_N^*}g_d(l) {\sf P}_{l,d}(t)\,K^{l}_{\rm eff}(r,r') \;,
\ee
with $K^{l}_{\rm eff}(r,r')$ given in Eq. (\ref{eq: K_eff_1d_supp}) and $g_d(l)$ is given in Eq. (\ref{eq: degen}) of the main text. 

Note that this formula for the kernel (\ref{kernel_finite_supp}), exact for any $N$, also holds for non-interacting fermions in an arbitrary spherically symmetric potential, i.e. with single particle Hamiltonian $H = -\q \Delta_{\bf x}+ V(|{\bf x}|)$. In this case, $K^{l}_{\rm eff}(r,r')$ is the kernel corresponding to $1d$ non-interacting fermions in an effective potential $V_{\rm eff}^l(r)$ as given in Eq. (\ref{Veff}) of the main text with the substitution $V_W(r) \to V(r)$ and $g_d(l)$ is given in Eq. (\ref{eq: degen}) of the main text. 

\subsubsection{Large $N$ analysis of the kernel at the edge}

We now analyse this formula in Eq. (\ref{kernel_finite_supp}) for large $N$, equivalently for large $k_F = \sqrt{2 m \, \mu/\hbar^2} \sim N^{1/d}$. As we will see, the sum over $l$ in Eq. (\ref{kernel_finite_supp}) is dominated by large values of $l$. Within each $l$-sector Eq. (\ref{kernel_finite_supp}) shows that the radial part $K_{\rm eff}^l(r,r')$ and the angular part ${\sf P}_{l,d}(t)$ are decoupled and we will thus analyze them separately in the limit of large $l$.

{\it Radial part}. We first analyze $K_{\rm eff}^l(r,r')$ given in Eq. (\ref{eq: K_eff_1d_supp}) in the limit of large $l$. We anticipate that the sum over $n$ is dominated by large values of $n$ and we thus determine the asymptotic behavior of $\chi_{n,l}(r)$ given in Eq. (\ref{chi_explicit_supp}) for both $l$ and $n$ large (and both of the same order $O(k_F)$ as we will see below). In this limit, we make use of the following asymptotic expansion of the Bessel function (the so called Debye's expansion) \cite{dlmf_supp}
\be\label{bessel_large_supp}
\J_{\nu}(k_{n,l}r)\sim\left(\frac{2}{\pi\sqrt{(k_{n,l}r)^2-l^2}}\right)^{\frac{1}{2}}\cos\left(k_{n,l}\xi\left(r,\frac{l}{k_{n,l}}\right)-\frac{\pi}{4}\right)\; {\rm with}\quad \xi(r,\tilde l)=\sqrt{r^2-\tilde l^2}-\tilde l\arccos\left(\frac{\tilde l}{r}\right) \;,
\ee
with $\nu = l + (d-2)/2 \sim l$ for large $l$. From this expansion (\ref{bessel_large_supp}), one can already obtain the expansion of $k_{n,l}$ for large $n$ and $l$. Indeed, by definition $k_{n,l}$ is the $n$-th real zeros of $J_{\nu}(x)$, i.e., $\J_{\nu}(k_{n,l}) = 0$. Hence from Eq. (\ref{bessel_large_supp}) with $r=1$ one obtains  
\begin{eqnarray}\label{condition_knl1}
k_{n,l} \, \xi\left(1,\frac{l}{k_{n,l}}\right) - \frac{\pi}{4} = \left(n + \frac{1}{2}\right) \pi \simeq n \, \pi \;\quad  {\rm for} \;\quad n \gg 1  \;.
\end{eqnarray}
Let us first apply this relation (\ref{condition_knl1}) to $n = m_l$ such that $k_{m_l,l} \simeq k_F$ [see Eq. (\ref{ml_supp})]. One obtains that for $l \gg 1$, $k_F \gg 1$, keeping $\tilde l = l/k_F$ fixed, $m_l$ takes the scaling form 
\be\label{m_l_supp}
m_l\sim k_F {\cal N}\left(\frac{l}{k_F}\right),\quad {\rm with}\quad {\cal N}(\tilde l)= \frac{1}{\pi}\xi(1,\tilde l) =  \frac{\sqrt{1-\tilde l^2}-\tilde l\arccos \tilde l }{\pi}.
\ee
One can easily check that ${\cal N}(\tilde l<1)>0$ and ${\cal N}(1)=0$ and therefore one concludes that the last occupied $l$-sector is such that $\tilde l = 1$, i.e. $l = l_N^* \approx k_F$. From Eq. (\ref{m_l_supp}) one sees that the typical scale of $l$ is $l = O(k_F)$ and Eq. (\ref{condition_knl1}) suggests that the typical scale of $n$ is also $k_F$. Furthermore, from (\ref{condition_knl1}) one obtains that for large $l$ and $n$, keeping $\tilde l = l/k_F$ and $\tilde n = n/k_F$ fixed, $k_{n,l}$ takes the scaling form
\begin{eqnarray}\label{knl_scaling}
k_{n,l} \approx k_F \, {\cal K}\left(\frac{n}{k_F},\frac{l}{k_F} \right) \;,
\end{eqnarray}  
where ${\cal K}(\tilde l, \tilde n)$ satisfies the equation (deduced easily from Eqs. (\ref{condition_knl1}) together with the expression of $\xi(1,\tilde l)$ in Eq. (\ref{bessel_large_supp}))
\begin{eqnarray}\label{eq_K}
\sqrt{{\cal K}^2 - \tilde l^2} - \tilde l \arccos\left(\frac{\tilde l}{{\cal K}} \right) = \tilde n \,\pi \;,
\end{eqnarray}
where we used the shorthand notation ${\cal K} \equiv {\cal K}(\tilde n,\tilde l)$. Note that, by definition of $m_l$ (\ref{ml_supp}), one has $k_{m_l,l} \approx k_F$. Since $m_l \approx k_F {\cal N}(\tilde l)$, the scaling form in Eq. (\ref{knl_scaling}) implies that ${\cal K}$ satisfies
\begin{eqnarray}\label{identity_K}
{\cal K}\left({\cal N}(\tilde l),\tilde l\right) = 1 \;.
\end{eqnarray}
Note also that, by differentiating Eq. (\ref{eq_K}) with respect to $\tilde n$, one obtains the identity
\begin{eqnarray}\label{deriv_K}
\frac{\partial {\cal K}}{\partial \tilde n}  = \frac{\pi \, {\cal K}}{\sqrt{{\cal K}^2- \tilde l ^2}} \;,
\end{eqnarray}
which will be useful in the following. 
 
We now use Eq. (\ref{bessel_large_supp}) to study the asymptotic form of the wave function $\chi_{n,l}(r)$ close to the wall at $r=1$.  A Taylor expansion near $r=1$ of the function $\xi(r,\tilde l)$ in this equation yields
\be\label{taylor_xi}
\xi(r,\tilde l)=\sqrt{1-\tilde l^2}-\tilde l\arccos (\tilde l)+(r-1)\sqrt{1-\tilde l^2}+O\left((1-r)^2\right)\,.
\ee
Inserting this expansion (\ref{taylor_xi}) into equation (\ref{bessel_large_supp}) and using the relation satisfied by $k_{n,l}$ in Eq. (\ref{condition_knl1}) one obtains for $(1-r) \sim k_F^{-1} \ll 1$
\begin{eqnarray}\label{inter1}
\J_{\nu}(k_{n,l}r)&\sim&\left(\frac{2}{\pi\sqrt{k_{n,l}^2-l^2}}\right)^{\frac{1}{2}}\cos\left( \left(n+\frac{1}{2}\right)\pi + (r-1)\sqrt{k_{n,l}^2 - l^2}) \right)  \nonumber \\
&\sim& \frac{(-1)^{n+1}}{\sqrt{k_F}} \left(\frac{2}{\pi\sqrt{{\cal K}^2-\tilde l^2}}\right)^{\frac{1}{2}} \sin{\left[k_F(r-1)\sqrt{{\cal K}^2 - \tilde l^2}\right]} \;.
\end{eqnarray} 
On the other hand, to compute the asymptotic behavior of $\chi_{n,l}(r)$ in Eq. (\ref{chi_explicit_supp}), we also need to analyze $\J_{\nu - 1}(k_{n,l}) = \J'_{\nu}(k_{n,l})$ (where we have used the relation $\J_{\nu-1}(x) = \J'_{\nu}(x) + (\nu/x) \J_{\nu}(x)$). From the asymptotic expansion in Eq. (\ref{bessel_large_supp}) one obtains
\begin{eqnarray}\label{inter2}
\J_{\nu-1}(k_{n,l}) = \J'_{\nu}(k_{n,l}) \sim \frac{1}{\sqrt{k_F}} (-1)^{n+1}  \left(\frac{2}{\pi\sqrt{{\cal K}^2-\tilde l^2}}\right)^{\frac{1}{2}} \, \frac{1}{{\cal K}}\sqrt{{\cal K}^2 - \tilde l^2}\;.
\end{eqnarray}
Therefore, using these asymptotic expansions (\ref{inter1}) and (\ref{inter2}) in the expression for $\chi_{n,l}(r)$ in Eq. (\ref{chi_explicit_supp}) one obtains that, for both $n$ and $l$ large, with $\tilde  n=n/k_F$ and $\tilde l = l/k_F$ fixed and $(1-r) = O(k_F^{-1})$, $\chi_{n,l}(r)$ behaves as
\be\label{expansion_chi}
\chi_{n,l}(r)\sim {\cal K}\sqrt{\frac{2}{{\cal K}^2-\tilde l^2}}\sin\left(k_F(r-1)\sqrt{{\cal K}^2-\tilde l^2}\right)\;,
\ee
where ${\cal K} \equiv {\cal K}(\tilde l,\tilde n)$ is given implicitly by the solution of equation (\ref{eq_K}). 

Using this asymptotic form (\ref{expansion_chi}), one can now compute the effective one-dimensional kernel
$K^{l}_{\rm eff}(r,r')$ for large $l = O(k_F)$ and $(r-1)$ as well as $(r'-1)$ of order $O(k_F^{-1})$. One obtains
\begin{eqnarray}\label{keff_1}
K^{l}_{\rm eff}(r,r') \sim \sum_{n=1}^{k_F {\cal N}(\tilde l)} \frac{2 {\cal K}^2}{{\cal K}^2 - \tilde l^2} \sin{\left(k_F(r-1)\sqrt{{\cal K}^2 - \tilde l^2}\right)} \sin{\left(k_F(r'-1)\sqrt{{\cal K}^2 - \tilde l^2}\right)}  \;.
\end{eqnarray}
In the limit of large $k_F = O(N^{1/d})$, the variable $\tilde n = n/k_F$ becomes continuous and the discrete sum of $n$ can be replaced by an integral
\begin{eqnarray}\label{keff_2}
K^{l}_{\rm eff}(r,r') \sim k_F \int_0^{{\cal N}(\tilde l)} d \tilde n \frac{2 {\cal K}^2}{{\cal K}^2 - \tilde l^2} \sin{\left(k_F(r-1)\sqrt{{\cal K}^2 - \tilde l^2}\right)} \sin{\left(k_F(r'-1)\sqrt{{\cal K}^2 - \tilde l^2}\right)} \;.
\end{eqnarray}
This integral can be performed explicitly by performing the change of variable $\tilde n \to z = \sqrt{{\cal K}(\tilde n, \tilde l)^2 - \tilde l^2}$. Indeed, using the identity in Eq. (\ref{deriv_K}) one has
\begin{eqnarray}\label{keff_3}
\frac{d z}{d \tilde n} = \frac{\partial {\cal K}}{\partial \tilde n} \frac{{\cal K}}{\sqrt{{\cal K}^2 - \tilde l^2}} = \pi \,\frac{{\cal K}^2}{{\cal K}^2 - \tilde l^2} \;.
\end{eqnarray}
Hence the integral in (\ref{keff_2}) can be written as
\begin{eqnarray}\label{keff_4}
K^{l}_{\rm eff}(r,r') \sim k_F \frac{2}{\pi} \int_0^{\sqrt{1 - \tilde l^2}} dz \, \sin[k_F(r-1)\, z] \sin[k_F(r'-1)\, z] \;.
\end{eqnarray}
where we have used $z({\cal N}(\tilde l))= \sqrt{{\cal K}({\cal N}(\tilde l),\tilde l) - \tilde l^2} = \sqrt{1 - \tilde l^2}$ [see Eq. (\ref{identity_K})]. Finally, performing explicitly the integral over $z$ one finds
\begin{eqnarray}\label{keff_5}
K^{l}_{\rm eff}(r,r') \sim k_F \sqrt{1 - \tilde l^2} K_1^{\rm e}\left(k_F \sqrt{1 - \tilde l^2}\,(r-1), k_F \sqrt{1 - \tilde l^2}\,(r'-1) \right) \;,
\end{eqnarray}
where we recall that
\begin{eqnarray}\label{k_edge_supp}
K_1^{\rm e}(a,b)=\frac{\sin(a-b)}{\pi(a-b)}-\frac{\sin(a+b)}{\pi(a+b)} \;.
\end{eqnarray}

{\it Angular part.} We now analyze the angular dependence of the kernel $K_\mu({\bf x}, {\bf y})$ in Eq.~ (\ref{kernel_finite_supp}), which within each $l$-sector is controlled by the function ${\sf P}_{l,d}(t)$, with $t=\frac{{\bf x}\cdot {\bf y}}{|{\bf x}|\,|{\bf y}|}=\cos\psi$, ${\psi}$ being the angle formed by the two vectors ${\bf x}$ and ${\bf y}$. Since we are interested in the limit where ${\bf x}$ and  ${\bf y}$ are close to each other, with $|{\bf x} - {\bf y}| = O(k_F^{-1})$, and also close to the boundary, i.e. $|{\bf x}|, |{\bf y}| \approx 1$, we are also interested in the regime where $\psi = O(k_F^{-1})$. From the differential equation satisfied by ${\sf P}_{l,d}(\cos\psi)$ in Eq. (\ref{sum_harm_supp}), one can show that when $l \gg 1$ and $\psi \ll 1$ keeping the product $l \, \psi$ fixed, it takes the scaling form 
\be
\frac{1}{S_d}{\sf P}_{l,d}(\cos\psi) \sim f_d(l\,\psi),\quad {\rm with}\quad f_d(u)=\frac{\Gamma(d-1)}{4\pi}\frac{\J_{\frac{d-3}{2}}(u)}{(4\pi u)^{\frac{d-3}{2}}}\;. \label{F_d}
\ee

{\it Scaling form of the full kernel.} With the help of these asymptotic forms both for the radial part (\ref{keff_5})-(\ref{k_edge_supp}) and for the angular part (\ref{F_d}) we can now analyze the asymptotic form of the kernel, in the large $k_F = O(N^{1/d})$ limit, and at the edge, i.e. close to the hard wall. In the edge scaling limit close to a boundary point ${\bf x}_w$, it is useful to parameterize the positions of the fermions as follows (see Fig. \ref{fig_image})
\bea\label{parameter}\label{def_an_at}
\begin{cases}
 &{\bf x}=\left(1+\dfrac{a_n}{k_F}\right){\bf x}_w+\dfrac{\bf{a}_t}{k_F}\\
 & \\
 &{\bf y}=\left(1+\dfrac{b_n}{k_F}\right){\bf x}_w+\dfrac{\bf{b}_t}{k_F}
\end{cases}
\eea
where ${\bf x}_w\cdot\bf{a}_t={\bf x}_w\cdot\bf{b}_t=0$. 
With this parameterization (\ref{parameter}), one can show that the radial kernel $K^{l}_{\rm eff}(|{\bf x}|,|{\bf y}|)$ only depends on $a_n$ and $b_n$ and takes the scaling form as in Eq. (\ref{k_edge_supp}), i.e.,
\begin{eqnarray}\label{scaling_an}
K^{l}_{\rm eff}(|{\bf x}|,|{\bf y}|) \approx  k_F \sqrt{1 - \tilde l^2} K_1^{\rm e}\left( \sqrt{1 - \tilde l^2} \, a_n, \sqrt{1-\tilde l^2} \, b_n\right) \;.
\end{eqnarray} 
On the other hand, the angular part ${\sf P}_{l,d}(\frac{{\bf x}\cdot{\bf y}}{|{\bf x}||{\bf y}|})$ only depends on $|{\bf a}_t-{\bf b}_t|$. And in the limit of large $l$, it takes the scaling form as in Eq. (\ref{F_d}), i.e.,  
\begin{eqnarray}\label{scaling_at}
\frac{1}{S_d}{\sf P}_{l,d}(\cos\psi) \approx f_d(\tilde l \, |{\bf a_t} - {\bf b_t}|) \;.
\end{eqnarray}
By injecting these scaling forms (\ref{scaling_an}) and (\ref{scaling_at}) in terms of these scaling variables (\ref{parameter}), into Eq. (\ref{kernel_finite_supp}), one obtains 
\begin{eqnarray}\label{Kmu_supp2}
K_{\mu}({\bf x}, {\bf y}) \sim k_F \, \sum_{l=0}^{l_N^*} g_d(l) \, f_d(\tilde l |{\bf a_t} - {\bf b_t}|) \sqrt{1 - \tilde l^2} K_1^{\rm e}\left( \sqrt{1 - \tilde l^2} \, a_n, \sqrt{1-\tilde l^2} \, b_n\right) \;.
\end{eqnarray} 
We recall that for large $k_F = O(N^{1/d})$, $l_N^* \approx k_F$ and the sum is dominated by the large values of $l = O(k_F)$ such that one can replace $g_d(l)$ by its large $l$ behavior 
\be
g_d(l) \approx \frac{2}{\Gamma(d-1)} l^{d-2} = k_F^{d-2} \frac{2}{\Gamma(d-1)} \tilde l^{d-2} \;. \label{gd_asympt} 
\ee

Furthermore, the scaled variable $\tilde l = l/k_F$ becomes continuous such that the discrete sum in Eq. (\ref{Kmu_supp2}) can be replaced by an integral over $\tilde l \in [0,1]$, yielding the scaling form
\be\label{kernel_ab}
K_{\mu}({\bf x},{\bf y}) \approx k_F^d K_{d}^{\rm e}\left(k_F({\bf x}-{\bf x}_w),k_F({\bf y}-{\bf x}_w)\right) \;,
\ee
where the scaling function $K_{d}^{\rm e}({\bf a},{\bf b})$ reads
\bea\label{kernel_ab2}
K_{d}^{\rm e}({\bf a},{\bf b}) = \frac{2}{\Gamma(d-1)} \int_0^1 d\tilde l \, \tilde l^{d-2} f_d(\tilde l |{\bf a_t} - {\bf b_t}|) \sqrt{1 - \tilde l^2} K_1^{\rm e}\left( \sqrt{1 - \tilde l^2} \, a_n, \sqrt{1-\tilde l^2} \, b_n\right) \;.
\eea
It turns out that the integral over $\tilde l$ can be explicitly computed. First we perform the natural change of variable $\tilde l = \sin \theta$ to obtain
\bea\label{kernel_ab3}
K_{d}^{\rm e}({\bf a},{\bf b}) = \frac{2}{\Gamma(d-1)} \int_0^{\pi/2} d\theta (\cos \theta)^2 \, (\sin \theta)^{d-2} f_d(|{\bf a_t} - {\bf b_t}|\sin \theta)  K_1^{\rm e}\left( a_n \, \cos \theta, b_n \cos{\theta} \right)
\eea
where $K_{1}^{\rm e}(a,b)$ and $f_d(u)$ are given respectively in Eqs. \eqref{k_edge_supp} and \eqref{F_d}. Quite remarkably, this integral over $\theta$ can be performed explicitly using the following identity \cite{dlmf_supp}
\be\label{integral_bessel}
\int_0^{\frac{\pi}{2}}(\cos\theta)^{\tau+1}\J_{\tau}(y\cos\theta)(\sin\theta)^{\sigma+1}\J_{\sigma}(z\sin\theta)d\theta
=\frac{z^{\sigma}y^{\tau}\J_{\tau+\sigma+1}(\sqrt{z^2+y^2})}{(z^2+y^2)^{\frac{\tau+\sigma+1}{2}}}\;.
\ee
Using this formula with $\tau=\frac{1}{2}$ and $\sigma=\frac{d-3}{2}$ to compute the integral in Eq. (\ref{kernel_ab3})
one obtains  
\be\label{k_d_final}
K_{d}^{\rm e}(a_n{\bf x}_w+{\bf a}_t,b_n{\bf x}_w+{\bf b}_t)=\frac{\J_{\frac{d}{2}}(\sqrt{(a_n-b_n)^2+({\bf a}_t-{\bf b}_t)^2})}{(2\pi \sqrt{(a_n-b_n)^2+({\bf a}_t-{\bf b}_t)^2})^{\frac{d}{2}}}-\frac{\J_{\frac{d}{2}}(\sqrt{(a_n+b_n)^2+({\bf a}_t-{\bf b}_t)^2})}{(2\pi\sqrt{(a_n+b_n)^2+({\bf a}_t-{\bf b}_t)^2})^{\frac{d}{2}}}\;.
\ee
Using finally that $\sqrt{(a_n-b_n)^2+({\bf a}_t-{\bf b}_t)^2} = |{\bf a} - {\bf b}|$ and $\sqrt{(a_n+b_n)^2+({\bf a}_t-{\bf b}_t)^2} = |{\bf a} - {\bf b}^T|$ (see Fig. \ref{fig_image}), Eq. (\ref{k_d_final}) gives the result announced in the Letter in Eq. (\ref{edge_explicit}) of the text.

\begin{figure}
\includegraphics[width=0.5\textwidth]{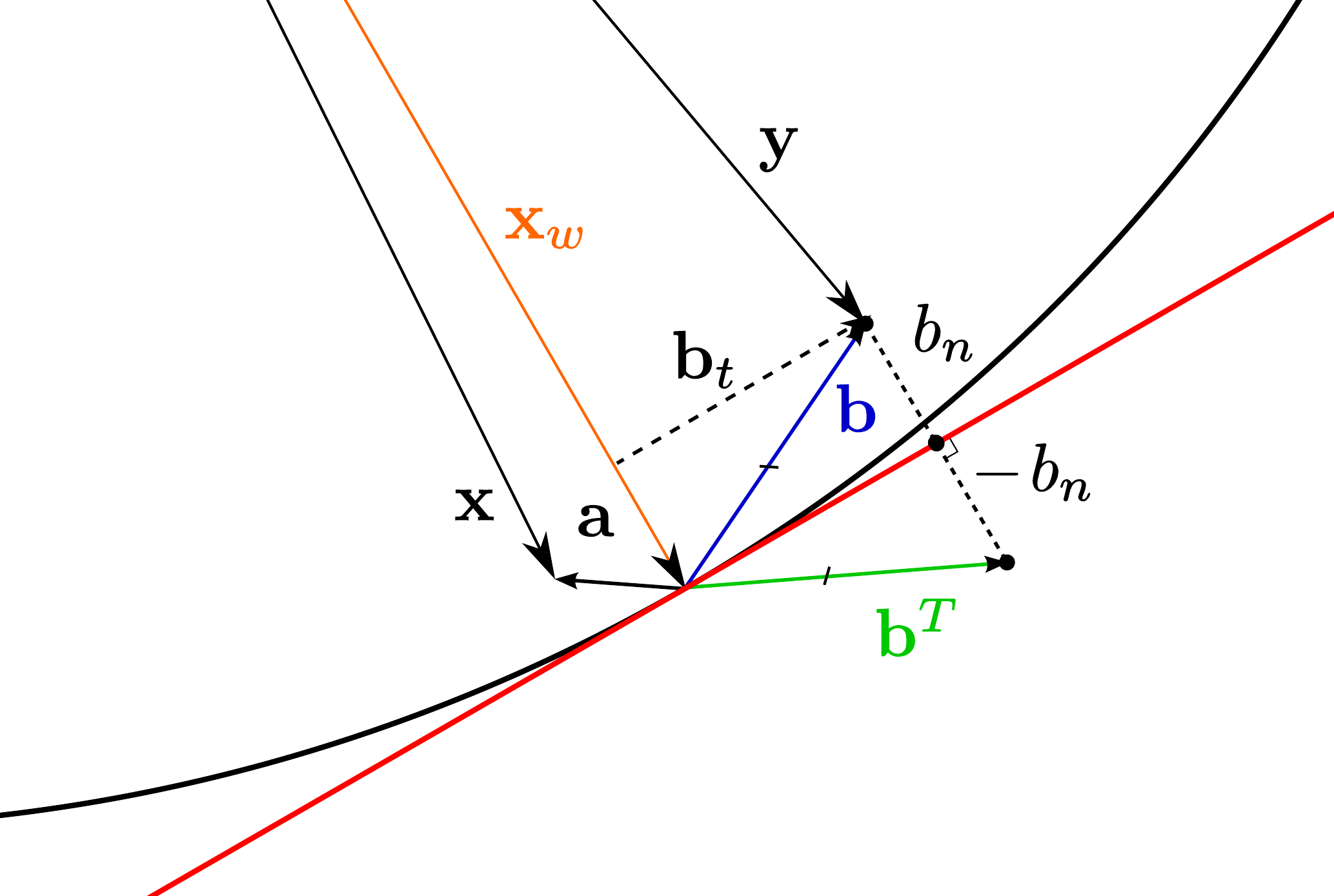}
\caption{Sketch of the method of images for the sphere (see Eqs. (\ref{def_an_at}) and (\ref{k_d_final})).}\label{fig_image}
\end{figure}

\subsubsection{Farthest fermion CDF $Q_d(w,N)$ in $d>1$}

In this section, we give some details regarding the results about the CDF $Q_d(w,N)$ of $r_{\max}$ in dimension $d>1$, for $N$ fermions in a spherically symmetric hard box, as in Eqs. (\ref{eq:HN_supp}), (\ref{def_VW_supp}).

{\it Intermediate deviation function.} Our starting point is the formula for $Q_d(w,N)$ given in Eq. (\ref{Qd_explicit_2}) of the main text (see also \cite{farthest_f_supp})
\be\label{Q_d_supp}
Q_d(w,N)=\exp\left(\sum_{l=0}^{l^*_N}g_d(l)\ln P_l(w,m_l)\right)\;,
\ee
where $P_l(w,m_l)$ is the CDF of the position of the rightmost fermion $x_{\max,l}$ among $m_l$ fermions (within each $l$-sector) in the $1d$ effective potential $V_{\rm eff}^l(r)$ given in Eq. (\ref{Veff}) in the main text. Now within each $l$-sector the $m_l$ fermions form a $1d$ determinantal process with an edge kernel, close to the boundary at $r=1$, given by Eq. (\ref{keff_5}). Therefore, up to the scale factor $\sqrt{1 - \tilde l^2}$, this is the same determinantal process as the one studied above [see Eq. (\ref{1d_k_supp}) and below]. One thus immediately concludes that, in the limit of large $k_F = O(N^{1/d})$, 
\be\label{Pl_1}
P_l(w,m_l) \approx \tilde{q}_1\left(k_F(1-w) \sqrt{1 - \tilde l^2}\right)\;, \quad \tilde l = \frac{l}{k_F} \;,
\ee
where we recall that $\tilde{q}_1(s)=\Det(I-P_s K_{1}^{\rm e}P_s)=D_{-}\left(\frac{2}{\pi}s\right)$ [see Eq. (\ref{eq: P_1d}) in the main text]. For later purpose, we also give its asymptotic behaviors~\cite{Grimm}
%
\be
\tilde q_1(s) =\left\lbrace\begin{array}{l}
1-\frac{2}{9\pi}s^3-\frac{2}{75\pi}s^5 + O(s^7),\ s\ll 1\\[0.2cm]
s^{-\frac{1}{8}} e^{-\frac{s^2}{4}+\frac{s}{2} + O(1)},\ s\gg 1
	\end{array}\right. .\label{eq: typ_fluc_1d_supp}
\ee
To analyse $Q_d(w,N)$ given in Eq. (\ref{Q_d_supp}) in the large $N$ (equivalently large $k_F = O(N^{1/d})$) limit we  replace $P_l(w,m_l)$ by its asymptotic form in Eq. (\ref{Pl_1}) as well as $g_d(l)$ by Eq. (\ref{gd_asympt}), as the sum over $l$ is dominated by large values of $l = O(k_F)$. Therefore, the discrete sum over $l$ can be replaced by an integral over $\tilde l = l/k_F$ (we recall that $\tilde l \in [0,1]$ since $l_N^* \approx k_F$). Hence in the large $N$ and large $l$ limit, one may write $Q_d(w,N)$ under the following scaling form
\be\label{Qd_supp1}
Q_d(w,N)\sim\exp\left(-k_F^{d-1}G_d(k_F(1-w))\right)\quad {\rm where}\quad G_d(s)=-\frac{2}{\Gamma(d-1)}\int_0^{1} \tilde l^{d-2} \ln \tilde{q}_1\left(s\sqrt{1-\tilde l^2}\right)d \tilde l\;,
\ee
which corresponds to the intermediate deviation regime, given in the second line of Eq. (\ref{summary_dgeq1}) in the main text. The asymptotic behavior of $G_d(s)$ can be simply obtained by replacing $\tilde{q}_1\left(s\sqrt{1-\tilde l^2}\right)$ by its appropriate asymptotic form, which can be read from Eq. (\ref{eq: typ_fluc_1d_supp}) and by performing the remaining integral over $\tilde l$. This yields  
\bea\label{Gd_asympt}
G_d(s) \approx
\begin{cases}
&(\alpha_d \, s)^3 \;, \; s \to 0 \; \quad {\rm with} \quad  \alpha_d=\left[3\times 2^{d-1}\Gamma\left(\frac{d}{2}+2\right)\Gamma\left(\frac{d}{2}\right)\right]^{-\frac{1}{3}}\\
& \\
& \frac{d}{(d+1)!} s^2 \;, \; s \to \infty \;,
\end{cases}
\eea
as announced below Eq. (21) of the main text.

{\it Typical region.} For large $k_F$, and for $d>1$, $Q_d(w,N)$ is non-zero only when $G_d(s) = O(k_F^{1-d}) \ll 1$, with $s = k_F(1-w)$, which happens when $s \to 0$ [see Eq. (\ref{Gd_asympt})]. By using the small $s$ behavior of $\tilde q_1(s)$ given in the first line of Eq. (\ref{eq: typ_fluc_1d_supp}) one can thus replace $\ln\tilde{q}_1(s\sqrt{1-\tilde l^2})$ by its asymptotic behavior for small $s$, $\ln\tilde{q}_1(s\sqrt{1-\tilde l^2})\sim -\frac{2}{9\pi}(1-\tilde l^2)^{\frac{3}{2}}s^3$ 
into the expression of $G_d(s)$ in Eq. (\ref{Qd_supp1}). Performing the remaining integral over $\tilde l$ we find
\be\label{Qd_supp2}
Q_d(w,N)\sim\exp\left(-[k_F^{\frac{d+2}{3}}\alpha_d (1-w)]^3\right)\;,\quad {\rm with}\quad \alpha_d=\left[3\times 2^{d-1}\Gamma\left(\frac{d}{2}+2\right)\Gamma\left(\frac{d}{2}\right)\right]^{-\frac{1}{3}} \;,
\ee
as announced in the first line of Eq. (\ref{summary_dgeq1}), which describes the typical behavior of $Q_d(w,N)$ (see also Fig. \ref{Fig_sketch} in the main text).

{\it Large deviation regime.} As in the $1d$ case [see Eq. (\ref{large_dev_1d_supp})], one expects that there is a large deviation regime associated to $Q(w,N)$ for $(1-w) = O(1)$. To unveil this regime, we recall that $P_l(w,m_l)$, within each $l$-sector, can be written as (see \cite{farthest_f_supp}) 
\begin{eqnarray}\label{P_explicit}
P_l(w,m_l) = \frac{1}{m_l!}\int_0^w dr_1 \cdots \int_0^w dr_{m_l} \left( \det_{1\leq i,j \leq m_l} \chi_{j,l}(r_i)\right)^2 \;,
\end{eqnarray}
where $\chi_{n,l}(r)$ is given in Eq. (\ref{chi_explicit_supp}). Computing this multiple integral (\ref{P_explicit}) for arbitrary $w$ with $1-w = O(1)$ seems very hard but progress can be made in the limit $w \to 0$. Indeed in this limit, one can replace the squared determinant in Eq. (\ref{P_explicit}) by its limiting behavior when $r_1, r_2, \cdots r_{m_l}$ are all small, i.e.,
\be\label{vdm_type}
\frac{1}{m_l!}\left(\det_{1\leq i,j\leq m_l}\chi_{j,l}(r_i)\right)^2\approx z_l \, \prod_{i=1}^{m_l} r_i^{2l+d-1}\prod_{i<j}^{m_l}\abs{r_i^2-r_j^2}^2 \;, \; \quad {\rm for} \quad r_1, r_2, r_{m_l} \ll 1 \;,
\ee
where $z_l$ is some constant, which is not important here. Inserting this asymptotic behavior (\ref{vdm_type}) into Eq. (\ref{P_explicit}) and performing the change variable $y_i = r_i/w$, the small $w$ behavior can the be simply obtained by power counting as
\be\label{final_Pl}
P_l(w,m_l) \approx z'_l w^{2 m_l(m_l+l+d/2-1)} \;.
\ee 
By inserting this small $w$ expansion (\ref{final_Pl}) into Eq. (\ref{Q_d_supp}) one obtains 
\be\label{Qd_supp2}
Q_d(w,N) \approx \exp(c_{d,N} \, \ln w) \;, \; w \to 0 \;\quad  {\rm with} \; \quad c_{d,N} = 2 \, \sum_{l=0}^{l_N^*} g_d(l)\,m_l \left(m_l + l +d/2-1 \right) \;.
\ee
This is the exact small $w$ behavior of $Q_d(w,N)$ for any finite $N$. In the large $N$ limit, the sum in (\ref{Qd_supp2}) is dominated by large values $l$ and one can replace $g_d(l)$ by its asymptotic behavior (\ref{gd_asympt}) and $m_l$ by its scaling form $m_l \approx k_F {\cal N}(\tilde l)$, with $\tilde l = l/k_F$ (\ref{m_l_supp}). For $k_F \gg 1$ the sum over $l$ can be replaced by an integral (we recall that $l_{N}^*\approx k_F$) and one finds
\be\label{Qd_supp3}
Q_d(w,N) \approx \exp{\left[k_F^{d+1} \, \kappa_d \ln w\right]} \;, \; \kappa_d = \frac{4}{\Gamma(d-1)} \int_0^1 d\tilde l \, \tilde l^{d-2} \, {\cal N}(\tilde l)({\cal N}(\tilde l) + \tilde l) \;.
\ee
This behavior (\ref{Qd_supp3}) is thus fully compatible with a large deviation form $Q_d(w,N) \approx \exp(-k_F^{d+1} \Phi_d(w))$, for $1-w = O(1)$, as given in the third line of Eq. (\ref{summary_dgeq1}) of the main text. In addition, the result in Eq. (\ref{Qd_supp3}) implies that $\Phi_d(w) \sim - \kappa_d \ln w$, as $w \to 0$, as announced in the paragraph above Eq. (\ref{eq: G_d}) in the main text. This integral over $\tilde l$ in Eq. (\ref{Qd_supp3}) can be computed explicitly in terms of hypergeometric functions. In particular one finds $\kappa_2 = 64/(27 \, \pi^2) = 0.2401\ldots$ or $\kappa_3 = 1/32 + 1/(2 \pi^2) = 0.0819 \ldots$.  

{\it Special case $d=1$.} For completeness, we mention that the large deviation function $\Phi_1(w)$ can be computed explicitly in $d=1$, along the lines exposed in section A.1 (relying on the mapping to the JUE). In that case, one finds $\Phi_1(w) = - \frac{4}{\pi^2} \ln \sin{\left(\frac{\pi w}{2} \right)}$, for $0<w\leq 1$. 


\section{B) Method of images and fermions in a wedge geometry}

Here we derive Eqs. \eqref{Kimages} in the text, we provide more details for the
argument leading to the method of images for a smooth boundary, and we derive
\eqref{wedgeKsmall} for the wedge. 

\subsection{B.1) Smooth boundary: method of images} 

For the box $]-R,R[$ in $d=1$ the standard method of images gives the propagator 
\be
G(x,y;t)= \sum_{n=-\infty}^{+\infty} \left(\frac{m}{2 \pi \hbar t}\right)^{\frac{1}{2}} \left( e^{- m \frac{(4 n R + x -y)^2}{2 \hbar t}} 
- e^{- m \frac{((4 n +2) R - x -y)^2}{2  \hbar t}}\right) \label{propag}
\ee
Inserting into \eqref{fromGtoK2} and using the identity valid for arbitrary $d$ and $a>0$
\be \label{integral} 
 \int_\Gamma \frac{dt}{2 i \pi t^{d/2+1}} \exp\left(z t - \frac{a}{t}\right) = \left(\frac{z}{a}\right)^{\frac{d}{4}} J_{\frac{d}{2}}(2 \sqrt{a z})
\ee 
we obtain the exact result, valid for any $\mu$ and $N$, for the kernel, \eqref{Kimages} in the text. 
We note that because of the decaying
behavior of the sine kernel, it is clear on Eq. (\ref{propag}) that (i) for $x,y$ inside the box and farther than 
$1/k_F \sim 1/N$ from either boundary walls only the direct term ($n=0$ of
first sum) contributes, leading to the standard sine kernel (ii) $x,y$ are within $1/k_F$ of
the wall at $x=R$, only the additional term $n=1$ of the second sum contributes (respectively only 
$n=-1$ near $x=-R$) (iii) all other images can be neglected, as they are much farther
than $1/k_F$ of any point in the box, i.e. the parameter $R \, k_F \gg 1$. 
As mentionned in the text, the latter feature is general, i.e. for large $\mu$ (see below) only nearby
images need to be considered. Note that the model immediately
extends to a rectangular box of size $R_x \times R_y$, in $d=2$ in the limit $R_y \gg R_x$, and
at fixed $\mu=\frac{\hbar k_F^2}{2 m}$, with $K_1^{\rm b} \to K_d^{\rm b}$ (in that
application $N$ is infinite if $R_y$ is taken infinite). 

Let us now detail the argument given in the text for $d=2$. Consider for
simplicity the model of the circular box of radius $R$. Let us shift the coordinates
for convenience so that the wall passes through $(0,0)$ and consider nearby points.
Inside the circle the euclidean propagator must satisfy $- \hbar \partial_t G=H G$, with $H
= \frac{\hbar^2}{2 m} (\partial_{y_1}^2 + \partial_{y_2}^2)$, and 
vanish on the wall, i.e. $G({\bf x},(y_1=f(y_2),y_2);t)=0$, 
where $y_1=f(y_2)=R \pm \sqrt{R^2-y_2^2}$ is the 
equation of the wall, with $f(y_2) \simeq y_2^2/(2 R)$ near $(0,0)$. 
Let us rewrite \eqref{fromGtoK2} denoting $t=\hbar \tilde t/\mu$ where $\tilde t$ is the
dimensionless rescaled time
\bea
&& K_\mu({\bf x}, {\bf y}) = \int_{\cal C} \frac{d\tilde t}{2 i \pi \tilde t} 
\exp(\tilde t) G({\bf x}, {\bf y};\hbar \tilde t/\mu)   = k_F^{d} \int_{\cal C} \frac{d\tilde t}{2 i \pi \tilde t} \exp(\tilde t) \tilde G(\tilde {\bf x}, \tilde {\bf y};
\tilde t)
\eea
where $G({\bf x}, {\bf y};\hbar \tilde t/\mu)   = k_F^{d} \tilde G(\tilde {\bf x}, \tilde {\bf y}; \tilde t)$.
In the second equality we use the dimensionless variables $\tilde {\bf x}= k_F {\bf x}$,  $\tilde {\bf y}= k_F {\bf y}$
where here we denote $\mu=\frac{\hbar^2}{2 m} k_F^2$. The dimensionless
propagator satisfies $-\partial_{\tilde t} \tilde G = (\partial_{\tilde y_1}^2 + \partial_{\tilde y_2}^2) \tilde G$
with $\tilde G(\tilde {\bf x}, \tilde {\bf y};0) = \delta^{d}(\tilde {\bf x}- \tilde {\bf y})$. 
The key point is now it must vanish 
$\tilde G({\bf x},(\tilde y_1=\tilde f(\tilde y_2),\tilde y_2);t)=0$
i.e. on the line $\tilde y_1=\tilde f(\tilde y_2)=k_F f(\tilde y_2/k_F) \simeq \tilde y_2^2/(2 k_F R) \approx 0$
using that $k_F R \gg 1$. Hence we see that only the shape of the wall near the considered
point $(0,0)$ matters (the remainder is sent to infinity) and that the effective radius of curvature of the
wall is now $k_F R$. Thus in the limit $k_F R \gg 1$ the wall can be considered as a plane
and the method of images applies in the region $\tilde {\bf x} \sim \tilde {\bf y} = O(1)$, i.e.
within distance $1/k_F$ from the wall. For the same reason as in the $1d$ discussion above, we
can focus on this region and neglect the contributions from other parts of the wall at distances
much larger than $1/k_F$. 

As discussed in the text the above argument can be extended to any finite domain ${\bf x} \in {\cal D}$ in any $d$ confined
by a smooth (twice differentiable) boundary $\partial {\cal D}$ acting as a hard wall, 
and in presence of an additional smooth potential $V({\bf x})$ inside the domain (until now
we have considered $V({\bf x})=0$). The control parameter of the problem is $\mu$,
the Fermi level energy, and one now defines $k_F({\bf x})= \sqrt{2 m(\mu-V({\bf x}))}/\hbar$.

{\it Bulk regime:} The small time expansion of \cite{fermions_review_supp} leads to a density in the bulk 
$\tilde \rho({\bf x})\simeq [k_F({\bf x})]^d/(2^d \gamma_d)$, as a function of $\mu$ for large $\mu$. This formula
is valid for ${\bf x}$ at distances $\gg 1/k_F({\bf x})$ from the wall. Note that this
result for the density in the bulk is independent of whether there is a wall or not. On the other hand the
total number of fermions is given by $\int_{{\bf x} \in {\cal D}} \tilde \rho({\bf x})=N$
and obviously depends on the presence of the wall.

{\it Hard wall edge regime:}  The conditions for the above arguments to be valid are as follows. First one must have at any boundary point ${\bf x}_w$,
$k_F({\bf x}_w) R({\bf x}_w) \gg 1$, where $R({\bf x}_w)$ has the following interpretation.
In dimension $d>2$ there are $d-1$ directions in the tangent plane to the wall at ${\bf x}_w$
each characterized by a radius of curvature. $R({\bf x}_w)$ is the minimum of all these
$d-1$ radii of curvatures. Then the method of images applies near the wall and the kernel is
given by $K_b^d$ minus its reflection with respect to the 
hyperplane tangent to the wall at that point as indicated in formula 
\eqref{edge_explicit}. If $V({\bf x}) \neq 0$ additional conditions are required,
namely that the potential does not vary too fast near the wall. This can be 
again obtained from the short time expansion performed in \cite{fermions_review_supp} 
(see Section VII there). To discard higher order terms in $t$, one needs that 
the characteristic time $t^*$ introduced in the text, which in presence of a potential renormalizes to $t^*=  \hbar/(\mu - V({\bf x}_w))$
be much smaller than $t_N$, the characteristic time for the smooth edge regime defined in Eq. (281) of Ref. \cite{fermions_review_supp}.
The condition $t^* \ll t_N$ leads to 
\be
\sqrt{\mu- V({\bf x}_{w})} \gg |V'({\bf x}_{w})|^{1/3} \hbar^{1/3} m^{-1/6} 
\ee
which can also be written as $k_F({\bf x}_{w}) \gg 1/w_N$, where $w_N$ is the
width of the smooth edge regime. If this condition is violated, one enters a more complicated edge regime in
presence of a hard wall, which we leave for future study. 

{\it Finite temperature.}
Following the same lines as \cite{fermions_review_supp} 
(see Section VII there) the same conclusions extend to finite temperature $T>0$
in the regime $T \sim \mu$ where $\mu$ is the Fermi energy defined in the paper.
Let us focus here on the case of the hard wall box with zero internal potential $V({\bf x})=0$.
One defines $\tilde \mu$ the finite temperature chemical potential of the grand canonical ensemble
as the solution of \cite{fermions_review_supp} 
\bea \label{mutilde} 
- {\rm Li}_{d/2}(- e^{\tilde \mu/T}) = \frac{N \lambda_T^d}{\Omega} 
\eea 
where ${\rm Li}_\nu(z) = \sum_{n\geq 1} \frac{z^n}{n^\nu}$, $\Omega$ is the volume of the box 
and $N$ the mean number of fermions in the box. We also introduce the de Broglie thermal wavelength
$\lambda_T=\hbar \sqrt{\frac{2 \pi}{m T}}$ and the "finite temperature Fermi momentum scale" 
$\tilde k = \sqrt{ 2 m \tilde \mu}/{\hbar}$. Using Eq. (240) of \cite{fermions_review_supp} and integrating by part we obtain the finite temperature
kernel $K^T_{\tilde \mu}({\bf x},{\bf y})$ as
\bea
&& K^T_{\tilde \mu}({\bf x},{\bf y}) = \int_{0}^{+\infty} d\mu' K_{\mu'}({\bf x},{\bf y}) \frac{1}{4 T [ \cosh \frac{1}{2 T} (\mu'-\tilde \mu)]^2} 
 \simeq \frac{\lambda_T^2}{8 \pi} \int_{0}^{+\infty} dk k^{d+1} K_d^{\rm e}(k {\bf x}, k {\bf y}) 
 \frac{1}{ [ \cosh \frac{\lambda_T^2}{8 \pi} (k^2-\tilde k^2)]^2} 
\eea
where in the last formula we have performed the change of variable $\mu'=\hbar^2 k^2/(2 m)$
and taken the large $N$ limit. For simplicity we have located the hard wall at ${\bf x}_w=0$.
Here $K_d^{\rm e}$ is the zero temperature edge kernel defined in the text 
in \eqref{result_edge}-\eqref{edge_explicit}, and given explicitly as
\bea
 && \label{edge2}
K_d^{\rm e}({\bf a},{\bf b}) = \frac{\J_{d/2}(|{\bf a}-{\bf b}|)}{(2\pi |{\bf a}-{\bf b}|)^{d/2}} - \frac{\J_{d/2}(|{\bf a}-{\bf b}^T|)}{(2\pi |{\bf a}-{\bf b}^T|)^{d/2} }
\eea 
where, as in the text, ${\bf b}^T$ is the
image of ${\bf b}$ by the reflection with respect to the tangent plane to the boundary at ${\bf x}_w=0$. 
When ${\bf x},{\bf y}$ are farther than a distance $\sim 1/k_F$ from the wall,
the second term in \eqref{edge2} vanishes and one recovers the finite temperature
bulk kernel given in an equivalent form in formula Eq. (274) in \cite{fermions_review_supp}
(note the misprint in the published version, see the correct formula (272) in arXiv version).
Note that as $k_F \lambda_T \gg 1$ one recovers the zero temperature result. 
Finally, the above analysis can be extended to an arbitrary internal smooth 
potential $V({\bf x})$ along the lines of \cite{fermions_review_supp}.

\subsection{B.2) Wedge geometry: method of images} 

Consider a wedge domain in $d=2$, with apex angle $\alpha$. Let us start
with the simplest case where $\alpha = \frac{2 \pi}{m}$ and integer $m$,
where the method of images, with multiple images, can be applied. Let us use complex plane 
coordinates $z=x_1+ i x_2$. Let us denote $\omega=e^{i \alpha}$ with
$\omega^m=1$, the rotations $R_j: z \to \omega^j z$ and reflections $T_j: z \to \omega^j \bar z$.
The method of images generates all compositions of the two reflections $T_0$ and $T_2$,
i.e. the set $R_{2k}$ and $T_{2k}$ for all $k \in \mathbb{Z}$. It is easy to see that
for even $m=2 p$ the resulting group is ${\cal G}: \{R_{2j},T_{2j} \}_{j=0,..p-1}$ with
$m$ elements, while for odd $m=2p+1$ odd it is ${\cal G}: \{R_{j},T_{j} \}_{j=0,..m-1}$
with $2 m$ elements. The resulting scaled kernel in the wedge reads 
\bea \label{kernelw1} 
&& K_2^{\rm e}(w,z) = \sum_{j=0}^{p-1} (K_2^{\rm b}(|w- \omega^{2j} z|) - K_2^{\rm b}(|w- \omega^{2j} \bar z|) \quad , \quad m=2 p \,\, \text{even} \\
&& K_2^{\rm e}(w,z) = \sum_{j=0}^{m-1} (K_2^{\rm b}(|w- \omega^{j} z|) - K_2^{\rm b}(|w- \omega^{j} \bar z|) \quad , \quad m=2 p+1 \,\, \text{odd} \;.
\eea 
The simplest example is the square $p=2$. The density at point ${\bf x}=(x_1,x_2)$ with $z=x_1+i x_2$, in the 
upper quadrant wedge $x_1,x_2>0$ is 
\bea
&& \tilde \rho(z) = \frac{k_F^2}{4 \pi} F_2^{\rm square}(k_F z) \label{densitysquare} \\
&& F_2^{\rm square}(z) = 4 \pi K_2^e(z,z) = 1 - \frac{J_1(2 x_1)}{x_1} - \frac{J_1(2 x_2)}{x_2} 
+ \frac{J_1(2 \sqrt{x_1^2+x_2^2})}{\sqrt{x_1^2+x_2^2}}   \simeq  \frac{1}{6} x_1^2 x_2^2 + O(x^6) 
\eea
where $J_1$ is the Bessel function. For the square quadrant, the density thus vanishes quartically with the distance to the apex. 

Until now the formula for the kernel \eqref{kernelw1} is exact for any $\mu$ for a wedge. Extensions
for a wedge made of smooth curved pieces, in presence of a smooth potential $V({\bf x})$ 
and at finite $T$ are immediate, along the lines of the previous paragraph, and with
similar conditions, since, again, it is a simple application of the method of images. 

\subsection{B.3) Wedge geometry : exact result for any angle} 

\subsubsection{Exact formula}

Here we work in units $m=\hbar=1$. We now use the exact formula given in
\cite{wedge1_supp} for the propagator in a wedge in $2d$ of apex angle $\alpha$.
Denoting $z=r e^{i \phi}$, $z_0=r_0 e^{i \phi_0}$, one has in polar coordinates
\bea 
G(z,z_0;t) &=& \frac{2}{\alpha} \sum_{n=1}^{+\infty} \sin\left(\frac{n \pi \phi}{\alpha}\right) 
\sin\left(\frac{n \pi \phi_0}{\alpha}\right) \int_0^{+\infty} dk \, k \, e^{- k^2 t/2} \J_{n \frac{\pi}{\alpha}}(k r) \J_{n \frac{\pi}{\alpha}}(k r_0) \label{Gwedge2} \\
& =& \frac{2}{\alpha t} \sum_{n=1}^{+\infty} \sin\left(\frac{n \pi \phi}{\alpha}\right) 
\sin(\frac{n \pi \phi_0}{\alpha}) I_{n \frac{\pi}{\alpha}}\left(\frac{r_0 r}{t}\right) \exp\left(- \frac{r^2 + r_0^2}{2 t}\right) \label{Gwedge} 
\eea 
which vanishes for $\phi=0,\alpha$. Here we have given two equivalent forms. We recall that the kernel $K_\mu (z,z_0)$ is related to
the propagator by the relation 
\be
\int_0^{+\infty} d\mu \, K_\mu (z,z_0) e^{- \mu t} = \frac{1}{t} G(z,z_0;t) \quad , \quad K_\mu(z,z_0) = \text{LT}^{-1}_{t \to \mu} ~ \frac{1}{t} G(z,z_0;t) \;.
\ee

{\it First form of the kernel.} Starting from the first form \eqref{Gwedge2} of the propagator, and using that $\text{LT}^{-1}_{t \to \mu} e^{- k^2 t/2}/t = \Theta(\mu-k^2/2)$
we can perform the integral over $k$ and obtain a first formula for the kernel in a universal form
\bea
K_{\mu}(z,z_0) = \frac{4 k_F^2}{\alpha} \sum_{n=1}^\infty \sin(\frac{n \pi \phi}{\alpha}) 
\sin(\frac{n \pi \phi_0}{\alpha})
K_{{\rm Bes},\frac{n\pi}{\alpha}}(k_F^2 r^2, k_F^2 r_0^2)
\eea
where 
\be
K_{{\rm Bes},\nu}(x,y) = \frac{1}{2(x-y)}( \J_\nu(\sqrt{x}) \sqrt{y} \J'_\nu(\sqrt{y}) - \J_\nu(\sqrt{y}) \sqrt{x} \J'_\nu(\sqrt{x})) \label{BesselK}
\ee
is the standard Bessel kernel known in RMT \cite{For10_supp}, and we replaced $\mu=k_F^2/2$ in our units. The density is thus
\bea
&&  \tilde \rho(z) = K_{\mu}(z,z) = \frac{k_F^2}{4 \pi} F_2^{{\rm wedge}, \alpha}(k_F z) \label{densitywedge2} \\
&& F_2^{{\rm wedge}, \alpha}(z) = \frac{4 \pi}{\alpha} \sum_{n=1}^\infty \sin^2\left(\frac{n \pi \phi}{\alpha}\right) 
\left( \J'_{ \frac{n \pi}{\alpha}}(r)^2 + \left(1 - \frac{n^2 \pi^2}{\alpha^2 r^2}\right) \J_{ \frac{n \pi}{\alpha}}(r)^2 \right) \label{densitywedge3} \;.
\eea
This series converges quickly and the formula is useful to plot the density (see Figure \ref{fig_wedge}).

\begin{figure}[ht]
\includegraphics[width = 0.6\linewidth]{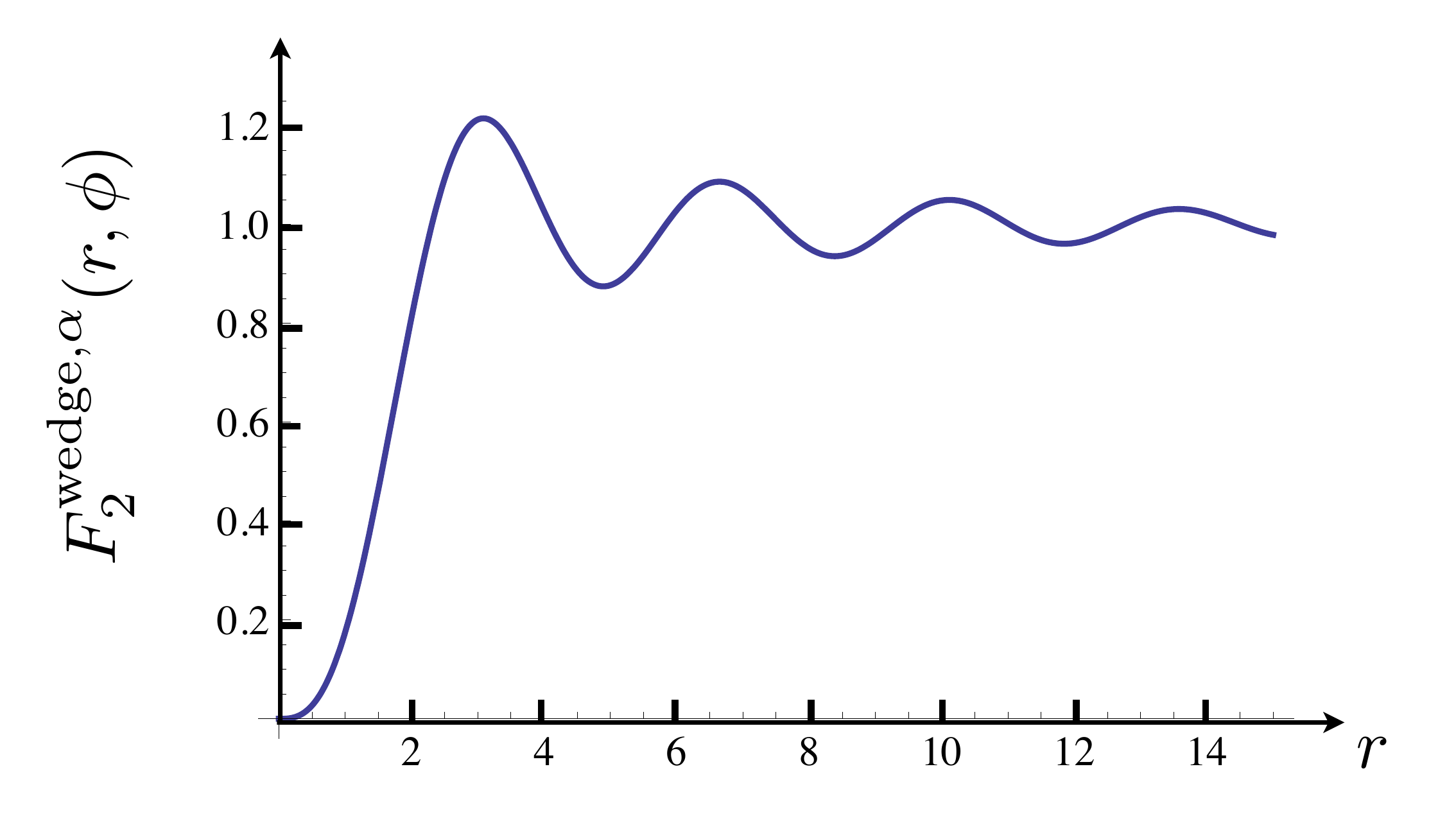}
\caption{Plot of $F_2^{{\rm wedge}, \alpha}(z) \equiv F_2^{{\rm wedge}, \alpha}(r,\phi)$ given in Eq. (\ref{densitywedge3}) as a function of $r$ for $\alpha = 7\pi/10$ and fixed $\phi = \alpha/2$.}\label{fig_wedge}
\end{figure}

{\it Second form of the kernel.} Let us give a formula which makes more apparent the
relation with the method of images. Let us now start from \eqref{Gwedge} and use the representation \cite{dlmf_supp}
\bea
I_\nu(y) = \frac{1}{\pi} \int_0^\pi d\psi \, e^{y \cos \psi} \cos(\nu \psi)  - \frac{\sin(\pi \nu)}{\pi} 
\int_0^{+\infty} du e^{-y \cosh u - \nu u} 
\eea
as well as formula \eqref{integral} to perform the inverse Laplace transform and obtain
\bea
&& K_\mu (z,z_0) =  \frac{2}{\alpha} \sum_{n=1}^{+\infty} \sin\left(\frac{n \pi \phi}{\alpha}\right) 
\sin\left(\frac{n \pi \phi_0}{\alpha}\right) \\
&& \times \bigg( \frac{1}{\pi} \int_0^\pi d\psi 
\cos(\frac{n \pi \psi}{\alpha}) \sqrt{\frac{2 \mu}{r^2+r_0^2- 2 r r_0 \cos(\psi)}} J_1(\sqrt{2 \mu (r^2+r_0^2- 2 r r_0 \cos(\psi))})
\\
&& -  \frac{\sin(n \frac{\pi^2}{\alpha})}{\pi} \int_0^{+\infty} du e^{- n \frac{\pi}{\alpha} u} 
\sqrt{\frac{2 \mu}{r^2+r_0^2+ 2 r r_0 \cosh(u)}} J_1(\sqrt{2 \mu (r^2+r_0^2+ 2 r r_0 \cosh(u))}) \bigg)
\eea 
The second term is absent when $\frac{\pi}{\alpha}=q$ with $q$ integer. Using the Poisson summation formula
$\sum_{n=-\infty}^{n=+\infty} e^{i n \theta} = 2 \pi \sum_{k=-\infty}^{k=+\infty} 
\delta(\theta- 2 \pi k)$, and using the constraints $0<\phi,\phi_0<\alpha$ and $0<\psi<\pi$,
leads a finite sum with alternate signs over images. In the general case, the summation over $n$
in both terms can also be achieved and leads to an explicit but complicated formula 
that we do not display here.

\subsubsection{Small distance expansion of the wedge kernel}

It is easy to perform the small distance expansion of the formula \eqref{Gwedge}. We use that
\bea
I_\nu(x) = \frac{x^\nu}{2^\nu \Gamma(1+\nu)} + O(x^{2 + \nu}) 
\eea
and obtain
\bea
G(z,z_0;t) \simeq \frac{2}{\alpha \, t 2^\frac{\pi}{\alpha}  \Gamma\left(1+ \frac{\pi}{\alpha} \right)} \sin\left(\frac{\pi \phi}{\alpha}\right) 
\sin\left(\frac{\pi \phi_0}{\alpha}\right) \left(\frac{r_0 r}{t}\right)^{\frac{\pi}{\alpha} }
\eea 
and we use that $LT^{-1}_{t \to \mu}  \frac{1}{t^{2 + \frac{\pi}{\alpha} }} = \frac{\mu^{1+  \frac{\pi}{\alpha}}}{\Gamma(2+  \frac{\pi}{\alpha})}$ to obtain the formula \eqref{wedgeKsmall} in the text. For the square wedge, $\alpha=\pi/2$, it agrees with the result \eqref{densitysquare}.


\subsubsection{General cone in $d$ dimensions}

One can consider a cone in dimension $d$, which is a direct generalization of a wedge in $d=2$. For any pair of points ${\bf x}$ and ${\bf y}$ belonging to the domain ${\cal W}$ bounded by the cone, the quantum propagator has the following expansion \cite{cone_supp}
\bea
 G({\bf x},{\bf y};t) &=& \sum_{j=1}^\infty m_j \left(\frac{{\bf x}}{|{\bf x}|} \right) m_j \left(\frac{{\bf y}}{|{\bf y}|}\right)
\int_0^{+\infty} dk \, k \, e^{-k^2 t/2} \frac{\J_{\alpha_j}(k |{\bf x}|)}{|{\bf x}|^{d/2-1}} \frac{\J_{\alpha_j}(k |{\bf y}|)}{|{\bf y}|^{d/2-1}}  \label{cone1} \\
&=& \frac{1}{t (|{\bf x}| |{\bf y}|)^{d/2-1}} \sum_{j=1}^\infty
m_j\left(\frac{{\bf x}}{|{\bf x}|}\right) m_j\left(\frac{{\bf y}}{|{\bf y}|})\right) I_{\alpha_j}\left(\frac{|{\bf x}||{\bf y}|}{t}\right) e^{- \frac{x^2+y^2}{2 t}} 
\quad , \quad  \alpha_j = \sqrt{\lambda_j + (d/2-1)^2}  
\eea
where $m_j(\bf x/|{x}|)$ are the eigenfunctions, with associated eigenvalues $\lambda_j$, of the Laplace-Beltrami operator (up to a minus sign), i.e. the operator ${\bf L}^2$, on ${\mathbb S}^{d-1}$ (unit sphere in ${\mathbb R}^d$)
with the condition that they vanish on the boundary of the domain ${\cal W}$. For the $2d$ cone $m_j(\phi)=\sqrt{\frac{2}{\alpha}} \sin \frac{\pi j \phi}{\alpha}$ and $\alpha_j=j \pi/\alpha$, and we recover \eqref{Gwedge}.  From Eq. \eqref{cone1} we derive, as in the previous section, 
\bea
&& K_\mu({\bf x},{\bf y}) =   \frac{2 k_F^2}{(|{\bf x}||{\bf y}|)^{d/2-1}} \sum_{j=1}^\infty m_j\left(\frac{{\bf x}}{|{\bf x}|}\right) m_j \left(\frac{{\bf y}}{|{\bf y}|}\right)
K_{{\rm Bes},\alpha_j}(k_F^2 |{\bf x}|^2, k_F^2 |{\bf y}|^2)
\eea
where $K_{{\rm Bes},\alpha_j}$ is the Bessel kernel \eqref{BesselK}, which arises from solving the radial
problem. From these expressions one can derive the small distance expansion of $K_\mu$
as
\bea
&& K_\mu({\bf x},{\bf y}) \simeq \frac{1}{2^{\alpha_1} \Gamma(1+\alpha_1)  \Gamma(2+\alpha_1)} 
(|{\bf x}||{\bf y}|)^{\sqrt{\lambda_1 + (d/2-1)^2} - (d/2-1)} 
m_1\left(\frac{{\bf x}}{|{\bf x}|}\right) m_1\left(\frac{{\bf y}}{|{\bf y}|}\right)  \quad , \quad \alpha_1 = \sqrt{\lambda_1 + (d/2-1)^2} \nn
\eea 
where $\lambda_1$ is the smallest eigenvalue of the Laplace-Beltrami operator (up to a minus sign) on the sphere
with vanishing conditions on the cone boundary. This generalizes the result \eqref{wedgeKsmall} given in the text
to a cone in any dimension $d$. 


{}

\end{widetext}

\end{document}